\documentclass[preprint,12pt]{elsarticle}
\usepackage{etoolbox}
\usepackage{comment}
\usepackage{pdflscape}
\usepackage{mathtools}
\makeatletter
\def\ps@pprintTitle{%
	\let\@oddhead\@empty
	\let\@evenhead\@empty
	\def\@oddfoot{\reset@font\hfil\thepage\hfil}
	\let\@evenfoot\@oddfoot
}
\makeatother

\usepackage{bm}
\usepackage{amsmath, amsthm, amssymb}
\usepackage{amsfonts}
\usepackage{natbib}
\usepackage{multicol}

\usepackage[colorlinks=true]{hyperref}
\usepackage{lscape}
\usepackage[utf8]{inputenc} 
\usepackage{amsthm}

\usepackage{enumerate}
\usepackage{mathrsfs}
\usepackage{tabularray}
\usepackage{placeins}
\usepackage{graphicx}
\usepackage{amsfonts}
\usepackage{verbatim}
\usepackage{amssymb}
\usepackage{amsthm}
\usepackage{multirow}
\usepackage{multicol}
\newtheorem{thm}{Theorem}[section]

\usepackage{xcolor}
\theoremstyle{definition}

\theoremstyle{remark}

\numberwithin{equation}{section}

\usepackage[mathscr]{euscript}
\usepackage{geometry} 
\usepackage{graphicx,lscape} 
\usepackage{float}
\usepackage{booktabs} 
\usepackage{bigstrut}
\usepackage{array} 
\usepackage{paralist} 
\usepackage{verbatim} 
\usepackage{subfig} 

\biboptions{comma,round,authoryear}

\usepackage{relsize}%
\usepackage{listings}
\usepackage[normalem]{ulem}
\useunder{\uline}{\ul}{}

\usepackage{natbib}

\usepackage{hyperref}
\hypersetup{
	colorlinks=true,
	allcolors=blue,
	pdftitle={A novel INAR(1) process based on the new McKenzie - error thinning operator with geometric marginals},
	pdfpagemode=FullScreen,
}

\usepackage{stackengine} 
\newcommand\oast{\stackMath\mathbin{\stackinset{c}{0ex}{c}{0ex}{\ast}{\bigcirc}}}

\usepackage[titletoc]{appendix}

\usepackage{enumitem}
\newlist{steps}{enumerate}{1}
\setlist[steps, 1]{label = Step \arabic*:}
\newlist{case}{enumerate}{1}
\setlist[case, 1]{label = Case (\roman*):}

\begin{document}
	
	\begin{frontmatter}
		
		\title{\textbf{Coherent forecasting of NoGeAR(1) model}}

			\author[label1]{Divya Kuttenchalil Andrews\corref{cor1}}
			\ead{divyaandrews5@gmail.com}
			\author[label1,label2]{N. Balakrishna}
			\cortext[cor1]{Corresponding author}
			\address[label1]{Cochin University of Science and Technology, Kochi, India.}
			\address[label2]{Indian Institute of Technology, Tirupati, India.}

		\begin{abstract}
			
			This article focuses on the coherent forecasting of the recently introduced novel geometric AR(1) (NoGeAR(1)) model - an INAR model based on inflated - parameter binomial thinning approach. Various techniques are available to achieve h - step ahead coherent forecasts of count time series, like median and mode forecasting. However,  there needs to be more body of literature addressing coherent forecasting in the context of overdispersed count time series. Here, we study the forecasting distribution corresponding to NoGeAR(1) process using the Monte Carlo (MC) approximation method. Accordingly, several forecasting measures are employed in the simulation study to facilitate a thorough comparison of the forecasting capability of NoGeAR(1) with other models. The methodology is also demonstrated using real-life data, specifically the data on CWß TeXpert downloads and Barbados COVID-19 data.\\
			
			\noindent MOS subject classification: 62M10\\
			
		\end{abstract}
		
	\end{frontmatter}

	
	\section{Introduction}
	\noindent Forecasting occupies a pivotal role in strategic decision-making and planning processes. The inherent uncertainty surrounding what lies ahead is both intriguing and demanding, prompting individuals and organizations to mitigate risks and optimize outcomes. Given the multitude of forecasting needs in various applications, there is a growing requirement for a varied range of forecasting methods to effectively address real-world challenges. Over the years, the proportion of publications concerning time series forecasting has exhibited a consistent level of stability. \cite{hyndman} and \cite{weiss21} provide comprehensive reviews of the progress of forecasting and count time series literature in the last few decades.\par
	
	However, with the emergence of count time series models, the need for obtaining integer - valued forecasts became relevant. In such cases, the traditional forecast, namely, the conditional mean, need not yield discrete - valued forecasts, thus leading to development of ``coherent" forecasting. The concept of coherent forecasting for Poisson integer autoregressive model (PINAR) was proposed by \cite{freeland2004forecasting}, using the median of the forecast distribution as a coherent forecast. The work also prompted research for other alternatives like Bayesian approach for point forecasts. (See 
	\cite{farrell2007hierarchical}, \cite{silva2009forecasting}, \cite{bisaglia2016bayesian} and \cite{homburg2019evaluating}). The coherent approach, however, favoured attention, as it required only the h-step ahead conditional distribution, which can be derived in most cases. Subsequently, \cite{jung2006coherent} and \cite{kim2010coherent}
	adopted the method for forecasting higher order INAR models. Throughout the remainder of the article, the h-step ahead ``conditional probability mass function" and ``forecasting distribution" may be used interchangeably.\par
	
	In recent times, much focus is centered on modeling and forecasting of overdispersed and zero - inflated count data. 
	Some of the  notable works on modeling overdispersed, underdispersed and zero-inflated count time series data include \cite{wang2001markov}, \cite{benjamin2003generalized}, \cite{weiss2009modelling}, \cite{zhu2012modeling}, \cite{maiti14}, \cite{bourguignon2017inar}, \cite{sathish22} and \cite{balki2023}. In the domain of forecasting overdispersed count time series data, \cite{maiti2015coherent} presented coherent forecasting of geometric INAR(GINAR) model. \cite{awale2017coherent} applied the procedure to get coherent forecasts of New geometric INAR (NGINAR) model and made comparisons with those of GINAR model. Later, \cite{awale2023forecasting} studied coherent forecasting for data modeled by negative binomial INAR(1) (NBINAR(1)) model. Contributions and applications in coherent forecasting were also made by \cite{maiti2015coherentb}, \cite{maiti2016coherent}, \cite{ristic2019zero}, \cite{guerrero2022integer} and \cite{khoo2022coherent}. Interestingly, studies on obtaining prediction intervals based on the forecasting distribution and using predictive likelihood for discrete-valued time series has been discussed in \cite{siuli} and \cite{weisspi}.\par
	
	In the present paper, we propose a coherent forecasting methodology for integer-valued time series data using NoGeAR(1) model. We define the two - step ahead conditional distribution using MC approximation. Simulation study is conducted to illustrate the efficacy of the proposed coherent forecasts within the framework of the NoGeAR(1) model, with comparative assessments against alternative INAR models. The application of our suggested methodology is exemplified through the analysis of two datasets, revealing close alignment between the forecasted values and actual outcomes when employing the NoGeAR(1) model for prediction.
	
	The structure of the paper is outlined as follows. In \textcolor{blue}{Section} \ref{constr}, we discuss the NoGeAR(1) model along with some basic properties. Coherent forecasting and measures of forecast accuracy are addressed in \textcolor{blue}{Section} \ref{forecast}. Simulation study is detailed in \textcolor{blue}{Section} \ref{sim} . The analysis of real datasets is presented in \textcolor{blue}{Section} \ref{data}. Finally, \textcolor{blue}{Section} \ref{conc} provides concluding remarks of the paper.

	\section{Brief introduction to NoGeAR(1) model}\label{constr}
	\noindent The NoGeAR(1) process $\{ X_t \}$, by \cite{doi:10.1080/00949655.2023.2213794}, is defined by
	\begin{equation}
		\label{eqn:eq1}
		X_{t} = \omega \text{\smaller[2]$ \oast$} X_{t-1} + \varepsilon_{t}, \quad t \geq 1,
	\end{equation}
	reprising the INAR(1) model assumptions - $\{X_t\}$ is a sequence of integer - valued random variables assuming non-negative values, $\omega \text{\smaller[2]$ \oast$}X_{t-1} = 0$  if  $X_{t-1}\equiv 0$, $0 \leq \omega < 1$, and $\{\varepsilon_t\}$ is the sequence of independent and identically distributed (i.i.d) innovations independent of the thinning operation and $X_{t-i}$ for all $t > i$. The distribution of $\omega\text{\smaller[2]$ \oast$} X_{t-1}$ given $X_{t-1} = j$ is specified by that of  $\sum _{i = 1} ^{j}G_{i} ^{\ast}$, where $\{G_i^{\ast}\}$ is a sequence of i.i.d random variables with, probability mass function (pmf):
	\begin{equation}
		\label{eqn:eq4}
		Pr[G^{\ast} = x]=\begin{cases} \alpha &\textrm{ if }  x=0,\\\\
			(1-\alpha)(1-\beta)\beta^{x-1} &\textrm{ if } x=1,2,\dots.
		\end{cases}
	\end{equation}
The thinning operator is denoted by `\text{\smaller[2]$ \oast$}' and $\omega$, the mean of $G^\ast$, equals $(1-\alpha)/(1-\beta)$. \cite{doi:10.1080/00949655.2023.2213794} established that when $0<\beta<\alpha<1$, the sequence $\{X_t\}$ defined by (\ref{eqn:eq4}) is stationary and each $X_t$ follows a geometric distribution with parameter $\theta$ ($0<\theta<1$) if and only if $\{\varepsilon_t\}$ follows a mixture of two geometric distributions with pmf:
	\begin{equation}
		\label{eqn:eq8}
		Pr[\varepsilon_t = x] =  \left( \frac{\alpha \theta - \beta}{\theta - \beta} \right) \theta ^x (1-\theta)  +  \left(
		1 - \frac{\alpha \theta - \beta}{\theta - \beta} \right) \beta ^x (1-\beta), \; x =0,1,2\dots.
	\end{equation}
	
	\noindent Under the above setup, for all non-negative integers $y,x $, the one-step transition probabilities are given by
	\small
	\begin{equation}
		\label{eqn: eq a}
		Pr[X_{t+1} = x| X_{t} = y] =\begin{cases}
			1 -\alpha \theta , & \textrm{if } x=0,\;y=0\\
			Pr[\varepsilon_t =x], & \textrm{if } x \geq 1,\;y=0\\
			\alpha ^y (1 -\alpha \theta) , & \textrm{if } x=0,\;y\geq 1\\
			\alpha ^y Pr[\varepsilon_t = x]\\
			+ \sum_{m = 1}^{x}\sum_{j = 1}^{min(m,y)} Pr[\varepsilon_t =x-m]\left(\begin{array}{l}
				y \\
				j
			\end{array}\right)\\   \times \left(\begin{array}{c}
				m-1 \\
				j-1
			\end{array}\right) [(1-\alpha)(1-\beta)]^j \alpha ^{y-j}\beta^{m-j}, &\textrm{if }x \geq 1,\;y\geq 1.
		\end{cases} 
	\end{equation} 
	\normalsize
	For more details on properties of the model and estimation of parameters, one may refer to \cite{doi:10.1080/00949655.2023.2213794}. In the following section, we discuss the concept of coherent forecasting and derive the two - step ahead conditional distribution.
	\section{Coherent forecasting}\label{forecast}
	
	\noindent After establishing the goodness of fit of a model for an observed INAR(1) process, a primary utility of the model then lies in forecasting future outcomes of the process. In the context of Box Jenkins' real-valued processes, the popular form of point forecast is typically the conditional mean, recognized for its optimality in minimizing the mean squared error. In count time series, however, the conditional mean may not be an integer, making it non-coherent. One approach to address this issue is to use a ceiling function to convert the forecasts into integers, thus ensuring coherence (See \cite{maiti2015coherent}).\par
	
	\cite{freeland2004forecasting} suggested alternative methods for achieving coherent forecasting by employing the conditional median or conditional mode, both of which consistently fall within the support space. Moreover, the median exhibits optimal characteristics, minimizing the prediction mean absolute error $E\{|X_{t+h} - \hat{X}_{t+h}|\big|\mathscr{F}_t\}$, for an $h$ - step ahead forecast, where $\mathscr{F}_t$ represents the history of the process upto time $t$. Considering the Markov nature of the INAR(1) process, $\mathscr{F}_t$ can be replaced by $X_t$. On the other hand, the mode represents the point where the predictive probability reaches its maximum value and is an optimal forecast according to the zero-one loss function, as discussed in \cite{gneiting2011making}. Note that if the conditional distribution shows multiple local modes, the global mode is to be considered for coherent forecasting. Based on simulated results discussed in \cite{maiti2015coherent}, it was observed that the median and mode predictors demonstrate significantly higher accuracy than the mean predictor when making h-step ahead forecasts for low count time series. One plausible explanation for this observation by \cite{awale2023forecasting} is that the forecasting distributions are skewed to the right and unimodal, leading to lower accuracy for the mean predictor compared to the median and mode predictors.\par
	
	The h-step ahead conditional mean and conditional variance of the process $\{X_t\}$ can be determined by conditioning over the intermediate random variables iteratively and using the one-step ahead conditional expectation in Equation (\ref{eqn:eq1}). That is,
	\begin{equation}
		E[X_{t+k}|X_t] = E\big[E[X_{t+k}\mid X_{t+k-1}]\big| X_t\big] \; \textrm{for} \; k=1,2,\ldots,h. \nonumber
	\end{equation}
	\cite{doi:10.1080/00949655.2023.2213794} have deduced the h - step ahead conditional mean and variance respectively as:
	\begin{equation}
		E[X_{t+h}|X_t] = \omega ^h X_t + \left( \frac{1-\omega ^h}{1-\omega}\right) \mu_\varepsilon,
	\end{equation}
	\noindent and
	\begin{equation}
		\begin{aligned}
			Var[X_{t+h}|X_t] &= \left(\frac{\alpha + \beta}{1-\beta}\right) \Bigg(\omega ^h \left(\frac{1-\omega ^h}{1-\omega}\right)X_t \\
			&\quad + \left(\frac{\omega(1-\omega^{2(h-1)})}{(1+\omega)(1-\omega)^2} - \frac{\omega^{2(h-1)}(1-\omega^{h-1})}{(1-\omega)^2}\right)\mu_\varepsilon\Bigg) + \left(\frac{1-\omega^{2h}}{1-\omega ^2}\right) \sigma_\varepsilon^2,
		\end{aligned}
	\end{equation}
	where, $\mu _\varepsilon$ and $\sigma^2_\varepsilon$ denote the common mean and common variance of $\{\varepsilon_t\}$ respectively. Nevertheless, to ascertain the median or mode, it is necessary to have the precise expression of the probability distribution. Consequently, the challenge lies in determining the h-step ahead forecasting distribution of the process described by (\ref{eqn:eq1}). The following subsection describes the commonly followed procedures to arrive at the h-step ahead conditional distribution, and the motivation for opting MC approximation to make coherent predictions.

	\subsection{The forecasting distribution}
	Using (\ref{eqn:eq1}), we have $X_{t+h} = \omega \, \text{\smaller[2]$ \oast$}\, X_{t+h-1} + \varepsilon_{t+h}, \, t \geq 1$. Now, by repeated substitution, one can arrive at:
	\begin{equation}
		X_{t+h} = \omega \, \text{\smaller[2]$ \oast$} \,\omega \,\text{\smaller[2]$ \oast$} \ldots \omega\, \text{\smaller[2]$ \oast$}\, X_t + \{\omega \,\text{\smaller[2]$ \oast$} \ldots \omega \,\text{\smaller[2]$ \oast$}\, \varepsilon _t + \ldots + \omega \,\text{\smaller[2]$ \oast$} \,\varepsilon_{t+h-1} + \varepsilon _{t+h}\},  
	\end{equation}
	which can be written as,
	\begin{equation}\nonumber
		X_{t+h} = \omega^{(h)}\text{\smaller[2]$ \oast$}X_t + \sum_{j=0}^{h-1} \omega^{(j)}\text{\smaller[2]$ \oast$}\varepsilon_{t+h-j},
	\end{equation}
	where $\omega ^{(h)}$ stands for operation of thinning $h$ times and $\omega^{(0)}=1$.
	So, we can formulate the h-step ahead conditional probability generating function (pgf), $P_{X_{t+h}|X_{t}}$, as
	\begin{equation}\label{eqaa1}
		P_{X_{t+h}|X_t}(s)= E[s^{X_{t+h}} |X_t] = \prod _{j=0}^{h-1}P_{\varepsilon}\Big(P_{G^\ast}^{(j)}(s)\Big)\Big[P_{G^\ast}^{(h)}(s)\Big]^{X_t},
	\end{equation}
	where $P_{G^{\ast}}^{(h)}(s) = P_{G^{\ast}}\left(P_{G^\ast}^{(h-1)}(s)\right)$, and $P_{G^{\ast}}^{(0)}(s) = s$. By (\ref{eqn:eq4}), we have 
		\begin{equation}
			P_{G^\ast}(s)=\frac{\alpha(1-s) + (1-\beta)s}{1-\beta s}.
		\end{equation}
		Then, for $h=1$, we have $P_{G^\ast}^{(1)}(s) = P_{G^\ast}(s)$. Subsequently, by mathematical induction, we can arrive at 
		\small
		\begin{equation}\label{pgfgstar}
			P_{G^{\ast}}^{(h)}(s) = \frac{\left((-1)^{h+1}\alpha^h+\sum_{i=1}^{h-1}(-1)^{i+1}\alpha^i\sum_{j=0}^{h-i}\binom{h}{j}(-\beta)^{h-j-i}\right)(1-s)+(1-\beta)^h s}{1+\beta(1-s)\sum_{i=1}^{h-1}(-\alpha)^i\sum_{j=0}^{h-i-1}\binom{h}{j}(-\beta)^{h-j-i-1} - \beta s \sum_{j=0}^{h-1}\binom{h}{j}(-\beta)^{h-j-1}}.
		\end{equation}
		\normalsize
		Equation (\ref{eqaa1}) was further simplified by \cite{doi:10.1080/00949655.2023.2213794} to:
	\begin{equation}\label{forecastpgf}
		P_{X_{t+h}|X_{t}}(s) =  P_{X}(s)\left[P_X\left(P_{G^\ast}^{(h)}(s)\right)\right]^{-1} \left[P_{G^\ast}^{(h)}(s)\right]^{X_t}.
	\end{equation}
	However, substituting (\ref{pgfgstar}) in (\ref{forecastpgf}) yields a complex form, which further complicates finding a closed form of pmf for h-step ahead conditional distribution. \cite{awale2017coherent} suggested a convolution approach to obtain the h-step ahead conditional pmf: 
	\begin{equation}
		\label{eq:for1}
		Pr[X_{t+h}=x|X_t = y] = \sum_{l=0}^xPr[\omega^{(h)}\text{\smaller[2]$ \oast$}y = l]Pr\left[\sum_{j=0}^{h-1}\omega ^{(j)}\text{\smaller[2]$ \oast$}\varepsilon_{t+h-j} = x-l\right].
	\end{equation}

	In particular, (\ref{pgfgstar}) has to be used for obtaining probabilities in (\ref{eq:for1}). The Markovian nature of $\{X_t\}$ implies that its state at time $t$ depends only on the state at $t-1$ and the higher order transition probabilities can be obtained using Chapman-Kolmgorov equations. In view of the complicated expressions of higher order transition probabilities, we provide the expressions for only one- and two- steps. The forecasts may be updated step by step when new observations are available. The two - step ahead transition probabilities can be computed as:
	\begin{equation}
		\label{eq:for2}
		Pr[X_{t+2}=x|X_t = y] = \sum_{k=0}^\infty Pr[X_{t+2}=x|X_{t+1} = k]Pr[X_{t+1}=k|X_t = y].
	\end{equation}
	The resulting form of the two - step ahead conditional pmf is presented in the following theorem:
	\begin{thm}
		\label{thm1}
		Let $\{X_t\}$ be a NoGeAR(1) process following (\ref{eqn:eq1}). Then, the two - step ahead transition probabilities are given by:
		\small
		\begin{equation}
			\label{eqn: eq a2}
			Pr[X_{t+2} = x| X_{t} = y] =\begin{cases}
				(1 -\alpha \theta)(1-\alpha \theta + \mathcal{A}), & \textrm{if } x=0,\;y=0\\
				\mathfrak{p}_{\varepsilon}(x)(1-\alpha \theta + \mathcal{A})+\sum_{k = 1}^{\infty}\mathfrak{p}_{\varepsilon}(k) \mathcal{B}(x,k), & \textrm{if } x \geq 1,\;y=0\\
				\alpha ^y (1 -\alpha \theta)\bigg(1 -\alpha \theta + \mathcal{A}+ \sum_{k = 1}^{\infty}\mathcal{B}(k,y)\bigg), & \textrm{if } x=0,\;y\geq 1\\
				\alpha ^y \mathfrak{p}_{\varepsilon}(x)\bigg(1 -\alpha \theta + \mathcal{A}+\sum_{k = 1}^{\infty}\mathcal{B}(k,y)\bigg)\\
				\quad+\alpha ^y \sum_{k = 1}^{\infty}\mathfrak{p}_{\varepsilon}(k)\mathcal{B}(x,k)\\
				\quad + \mathfrak{p}_{\varepsilon}(x)\sum_{k = 1}^{\infty}\mathcal{B}(x,k)\mathcal{B}(k,y), &\textrm{if }x \geq 1,\;y\geq 1,
			\end{cases} 
		\end{equation} 
		where, $\mathfrak{p}_{\varepsilon}(x)\coloneqq Pr[\varepsilon_t =x]$,  $\mathcal{B}(x,k)=\sum_{m = 1}^{x}\sum_{j = 1}^{min(m,y)} \left(\begin{array}{l}
			y \\
			j
		\end{array}\right)\left(\begin{array}{c}
			m-1 \\
			j-1
		\end{array}\right) \times [(1-\alpha)(1-\beta)]^j \alpha ^{y-j}\beta^{m-j}\mathfrak{p}_{\varepsilon}(x-m)$, and $\mathcal{A}=\sum_{k=1}^{\infty}\alpha^k\mathfrak{p}_{\varepsilon}(k)$.
	\end{thm}
	A proof of \textcolor{blue}{Theorem} \ref{thm1} is provided in \textcolor{blue}{Appendix} \ref{appA}. It may be observed that the value of $k$ in the summation terms range up to infinity due to the support of the distribution constituting non-negative integers. Also, the above exercise could be more tedious when extended to higher order forecasting distributions.   \par
		Considering such instances, wherein a closed form of the pmf is not available, \cite{weiss2018introduction} proposed using the MC approximation. That is, one can define the matrix $\widetilde{\mathbb{P}} \coloneqq (\mathfrak{p}_{yx}) $, $y,x = 0,1,\dots \mathcal{M}$, where $\mathcal{M}$ is a sufficiently large positive integer and $\mathfrak{p}_{yx}$ denotes the transition probabilities given by (\ref{eqn: eq a}). Then, by virtue of the Markov property, the h- step ahead transition probabilities, $\mathfrak{p}^{[h]}_{yx}$, may be obtained from the approximated matrix $\widetilde{\mathbb{P}}^{h} \coloneqq (\mathfrak{p}^{[h]}_{yx}) $, $y,x = 0,1,\dots \mathcal{M}$. Thus, for practical purposes, one can obtain the MC approximated form of two-step ahead transition probabilities by simply restricting the upper limit of $k$ in the summations in (\ref{eqn: eq a2}) to $\mathcal{M}=200$, say. Moreover, considering higher computation time required for higher order forecasting probabilities, the present paper confines to two - step ahead coherent forecasting of NoGeAR(1). 
	\begin{figure}[H]
		\centering
		\includegraphics[scale=0.5]{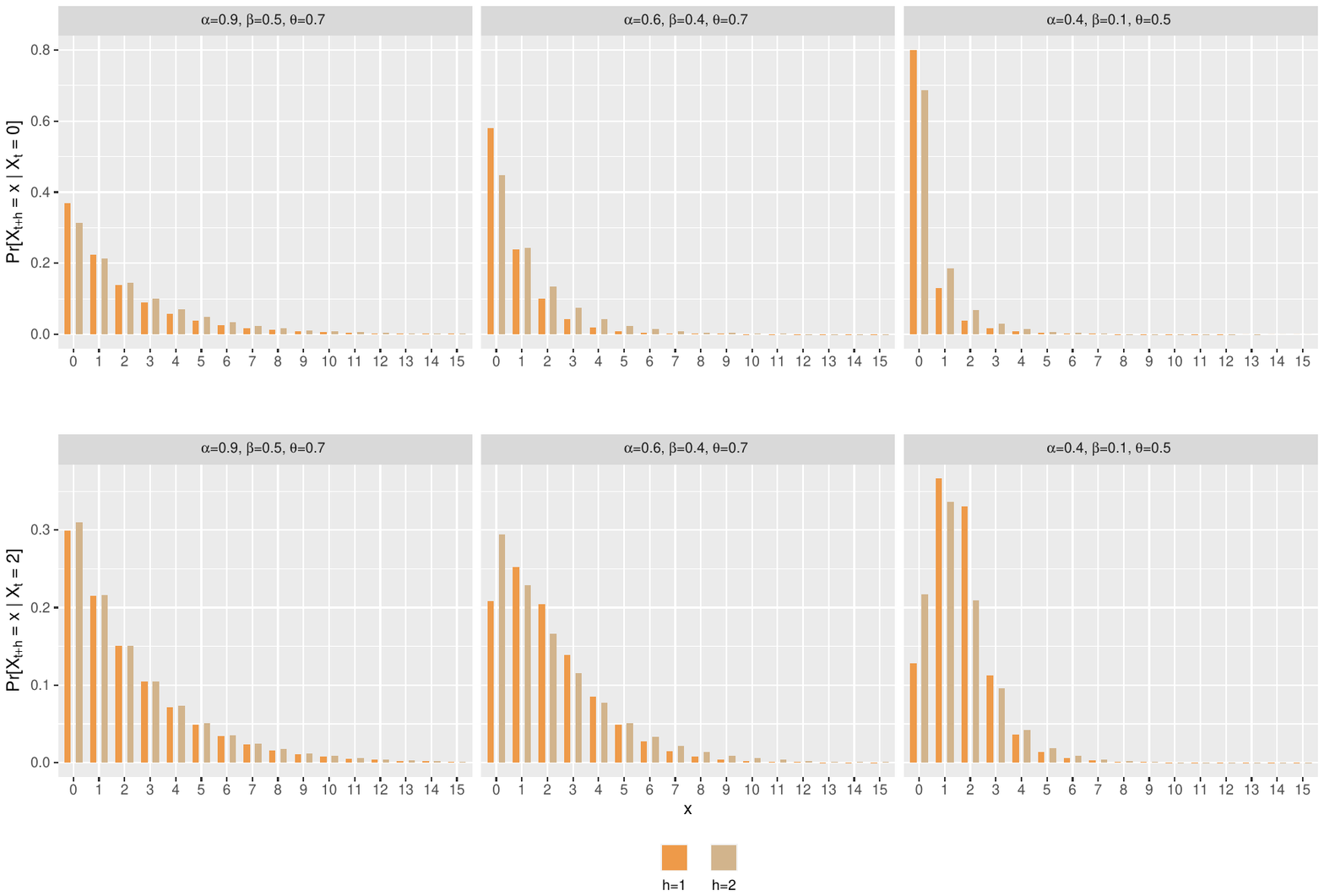}
		\caption{Plots of forecasting distributions for horizons, $h=1$ and $h=2$ for various combinations of parameters of NoGeAR(1) model given the values $X_t=0$ and $X_t=2$. }
		\label{fig:1}
	\end{figure}
	
	\subsection{The highest predictive probability (HPP) interval}
	
	\textcolor{blue}{Figure} \ref{fig:1} depicts plots of one-step and two-step ahead forecasting distributions for various parameter combinations. It is evident that the forecast distribution is characterized by significant skewness and peakedness across all considered parameter combinations. Consequently, it is clear that attempting to approximate this distribution with a Gaussian distribution would not be recommendable. Likewise, the use of standard prediction intervals which assume the predictive probability distribution to be symmetric cannot be applied in this context. For this, \cite{wang2021some} considered the highest predictive probability (HPP) interval as a solution. According to their method, the $100(1-\delta)\%$ HPP interval, of $X_{t+h}$ given $X_t$, is defined as $\mathcal{C}^{[h]} \equiv (X_{\mathscr{L}},X_{\mathscr{U}}) \coloneqq \{x:\mathfrak{p}^{[h]}_{xy} \geq \mathcal{K}_{\delta}\}$, where $\mathcal{K}_{\delta}\}$ is the largest number satisfying
	\begin{equation}\nonumber
		Pr[X_{\mathscr{L}}\leq X_{t+h} \leq X_{\mathscr{U}}|X_t=y] = \sum_{x=X_{\mathscr{L}}}^{X_{\mathscr{U}}}\mathfrak{p}^{[h]}_{xy} \geq 1-\delta.
	\end{equation}
	The computation of HPP intervals is made possible by the algorithm detailed in \cite{wang2021some} and \cite{awale2023forecasting}. The algorithm is not detailed in this paper to avoid repetition. \textcolor{blue}{Table} \ref{sim1} contains the coverage probabilities of $95\%$ HPP intervals for $h=1,2,$ corresponding to four parameter combinations and for samples sized $100$, $300$, $500$ and $1000$. From \textcolor{blue}{Table} \ref{sim1}, it can be observed that the
	
	\begin{table}[H]
		\center
		\caption{Coverage probabilities of $95\%$ HPP intervals  for various parameter configurations}
		\renewcommand{\arraystretch}{1.5}
		\scalebox{0.77}{%
			\begin{tabular}{lcccclcccc}
				\hline
				\multicolumn{1}{c}{} & \multicolumn{4}{c}{\textbf{(I) $\alpha=0.6$, $\beta=0.4$, $\theta =0.75$}}                                                                                  &  & \multicolumn{4}{c}{\textbf{(III) $\alpha=0.55$, $\beta=0.45$, $\theta=0.83$}}                                                                                 \\ \cline{1-5} \cline{7-10} 
				& \multicolumn{1}{l}{\textbf{$n=100$}} & \multicolumn{1}{l}{\textbf{$n=300$}} & \multicolumn{1}{l}{\textbf{$n=500$}} & \multicolumn{1}{l}{\textbf{$n=1000$}} &  & \multicolumn{1}{l}{\textbf{$n=100$}} & \multicolumn{1}{l}{\textbf{$n=300$}} & \multicolumn{1}{l}{\textbf{$n=500$}} & \multicolumn{1}{l}{\textbf{$n=1000$}} \\ \cline{2-5} \cline{7-10} 
				\textbf{h=1}         & 0.9048                               & 0.9132                               & 0.9379                               & 0.9437                                &  & 0.8866                               & 0.9058                               & 0.9124                               & 0.9200                                \\
				\textbf{h=2}         & 0.9118                               & 0.9372                               & 0.9398                               & 0.9477                                &  & 0.8915                               & 0.9155                               & 0.9325                               & 0.9480                                \\ \cline{1-5} \cline{7-10} 
				\multicolumn{1}{c}{} & \multicolumn{4}{c}{\textbf{(II) $\alpha=0.7$, $\beta = 0.3$, $\theta = 0.5$}}                                                                              &  & \multicolumn{4}{c}{\textbf{(IV) $\alpha=0.8$. $\beta=0.2$, $\theta =0.5$}}                                                                                 \\ \cline{1-5} \cline{7-10} 
				& \multicolumn{1}{l}{\textbf{$n=100$}} & \multicolumn{1}{l}{\textbf{$n=300$}} & \multicolumn{1}{l}{\textbf{$n=500$}} & \multicolumn{1}{l}{\textbf{$n=1000$}} &  & \multicolumn{1}{l}{\textbf{$n=100$}} & \multicolumn{1}{l}{\textbf{$n=300$}} & \multicolumn{1}{l}{\textbf{$n=500$}} & \multicolumn{1}{l}{\textbf{$n=1000$}} \\ \cline{2-5} \cline{7-10} 
				\textbf{h=1}         & 0.7835                               & 0.8808                               & 0.9213                               & 0.9307                                &  & 0.8770                               & 0.8916                               & 0.9053                               & 0.9189                                \\
				\textbf{h=2}         & 0.8523                               & 0.9109                               & 0.9305                               & 0.9402                                &  & 0.8793                               & 0.9314                               & 0.9437                               & 0.9463                                \\ \hline
		\end{tabular}}
		\label{sim1}
	\end{table}
\noindent coverage probabilities converge to 0.95 in most cases for $h=2$. Evidently, for $h=1$, the convergence is slow, possibly due to the skewness of forecast distribution in (\ref{eqn: eq a}). The next section entails the simulation study conducted to analyse the performance of the coherent forecasts of the NoGeAR(1) model as compared to other INAR models.
	\section{Simulation study}\label{sim}
	In this section, some simulation experiments are conducted to facilitate a comparison between NoGeAR(1) model and several other INAR(1) models, viz., the NGINAR model (\cite{ristic2009new}), GINAR model (\cite{mckenzie1986autoregressive}) and PINAR model (\cite{al1987first}). Additionally, to assess the robustness of NoGeAR(1) process, we generate data from NGINAR model as well. For the purpose of comparing models,  three metrics of forecasting accuracy have been examined, which will be defined in the subsequent subsection.
	
	\subsection{Measures of forecasting accuracy}
	Consider a set of $n+m$ observations, $\{X_1, X_2, \ldots, X_n, X_{n+1},\ldots, X_{n+m}\}$,  sampled from an INAR(1) model. Suppose the initial $n$ observations are treated as `training set' and utilized for model fitting, and the remaining $m$ observations, the `test set', are employed to validate the fitted model through one and two - step ahead forecasting. To assess the forecast accuracy of the model, the commonly used measures include - the prediction root mean squared error (PRMSE), the prediction mean absolute deviation (PMAD), and the percentage of true
	prediction (PTP). These measures are defined as follows:
	\begin{enumerate}
		\item[(i)] prediction root mean squared error (PRMSE)\small\begin{equation}\nonumber
			PRMSE(h) = \sqrt{\frac{1}{N} \sum_{i=1}^{N}\left( X_{(t_i)} - hat{X}^{(h)\text{Mean}}_{(t_i-h)}\right)^2};   t_i=(n+1)_i, (n+2)_i,\dots (n+m)_i; \, h=1,2,
		\end{equation}\normalsize
		where $N$ is the number of iterations, and $\hat{X}^{(h)\text{Mean}}_{(t_i-h)}$ is the h-step ahead conditional mean given $ X_{(t_i)}$ (the $t^{th}$ observation of the $i^{th}$ simulated sample) rounded to the nearest integer. For values of $X_{t_i -h}$ belonging to `test set', the one - step and two-step ahead forecasts from the corresponding previous time points are considered respectively. 
		\item[(ii)] prediction mean absolute deviation (PMAD)\begin{equation}\nonumber
			PMAD(h) = \frac{1}{N} \sum_{i=1}^{N}\left|X_{(t_i)} - \hat{X}^{(h)\text{Med}}_{(t_i-h)}\right|; t_i=(n+1)_i, (n+2)_i,\dots (n+m)_i; \, h=1,2,
		\end{equation}
		where $N$ reprises the earlier definition and $\hat{X}^{(h)\text{Med}}_{(t_i-h)}$ is the h-step ahead conditional median given $ X_{(t_i)}$. 
		\item[(iii)] percentage of true
		prediction (PTP)\begin{equation}\nonumber
			PTP(h) = \frac{1}{N} \sum_{i=1}^{N} \mathcal{I}(X_{(t_i)} = \hat{X}^{(h)}_{(t_i-h)}) \times 100\%,
		\end{equation}
		where $\mathcal{I}(.)$ denotes the indicator function and $\hat{X}^{(h)}_{(t_i-h)}$ can be any kind of point forecast - conditional mean, conditional median or conditional mode  (See \cite{khoo2022coherent}).
	\end{enumerate}
	The details of the data generating process used to examine the relative performances of the aforementioned models are now provided. As mentioned, two cases, viz., when data generating process is NoGeAR(1) (Case (i)) and when data generating process is NGINAR (Case (ii)) have been considered. The effect of sample size on forecasting performances has been explored by considering two different sample sizes - a small sample of  $n + m = 200$, and a moderate sample of $n + m = 500$. The training sets consist of $70\%$ of the simulated data and $30\%$ were reserved for the test set. The estimation of parameters using the training sets were done using Saddlepoint Approximation method (See \cite{doi:10.1080/00949655.2023.2213794}). Conducting these simulations was time-intensive due to the generation of higher counts for series with large parameter values, resulting in extended computation times. To manage this, the analysis was limited to 100 repetitions of a series for a particular parameter combination and sample size. \\ \\
	\noindent \textit{Case (i) Data generated from NoGeAR(1) process}\\ \par
	The first simulation study was performed for data generated from NoGeAR(1) model. Based on the stationarity conditions of the models considered, four parameter configurations were chosen - (I) $\alpha=0.8$, $\beta=0.2$, $\theta=0.5$, (II)
	$\alpha=0.6$, $\beta = 0.4$, $\theta = 0.75$,
	(III) $\alpha=0.7$, $\beta = 0.3$, $\theta = 0.5$, and
	(IV) $\alpha=0.55$. $\beta=0.45$, $\theta =0.83$. \textcolor{blue}{Table} \ref{tab2} represents the average forecast accuracy measures obtained for NoGeAR(1) and NGINAR models when data are from NoGeAR(1) model with the minimum values emphasized in bold. As observed, the forecasting measures generally increase with h across most models, suggesting that as we project further into the future, the forecasted values diverge from the actual observations. NoGeAR(1) and NGINAR models exhibit the lowest forecasting errors in terms of PRMSE and PMAD, indicating the higher forecasting accuracy. This outcome is expected since the true data generating process is NoGeAR(1). \par
	In \textcolor{blue}{Table} \ref{tab3}, the PTP of forecasts obtained using NoGeAR(1) and NGINAR models when the data are generated from NoGeAR(1) model are presented. It is observed that the PTPs are higher for mode and median forecasts. In general, the PTPs are shown to be consistent with increase in sample size.\\
	
	\noindent \textit{Case (ii) Data generated from NGINAR(1) process}\\ \par
	
	\textcolor{blue}{Table} \ref{tab4} displays the average PRMSE and PMAE for the models when data are drawn from NGINAR(1) model wherein the minimum values are given in bold. In this case too, the parameters of NGINAR model are chosen considering the stationary conditions of the models under study. Interstingly, in some cases, the NoGeAR(1) model outperforms other INAR(1) models in terms of PMAD and at times even PRMSE. The superior performance of NoGeAR(1) over NGINAR in those cases is attributed  to the fact that when $\alpha = 1 - \beta = \frac{1}{1+\alpha^{ng}}$ (`$\alpha^{ng}$' represents the parameter `$\alpha$' of NGINAR model to avoid confusion), NoGeAR(1) simplifies to NGINAR(1). With regard to the median forecast, as assessed by PMAD, the GINAR model also demonstrates competitiveness in this aspect, similarly to the NoGeAR(1) model. \textcolor{blue}{Table} \ref{tab5} presents the PTP values for the INAR(1) models when data are generated using NGINAR(1) model and conclusions analogous to \textcolor{blue}{Table} \ref{tab3} can be affirmed here as well. \par
	Overall, the forecast accuracy measures for NoGeAR(1) model were found to be smaller and hence better as compared to those of NGINAR(1), even when the data are from NGINAR(1) model. 
	\section{Data Analysis}\label{data}
	\subsection{Daily cases of COVID-19 in Barbados}
\begin{figure}[H]
		\centering
		\includegraphics[scale=0.5]{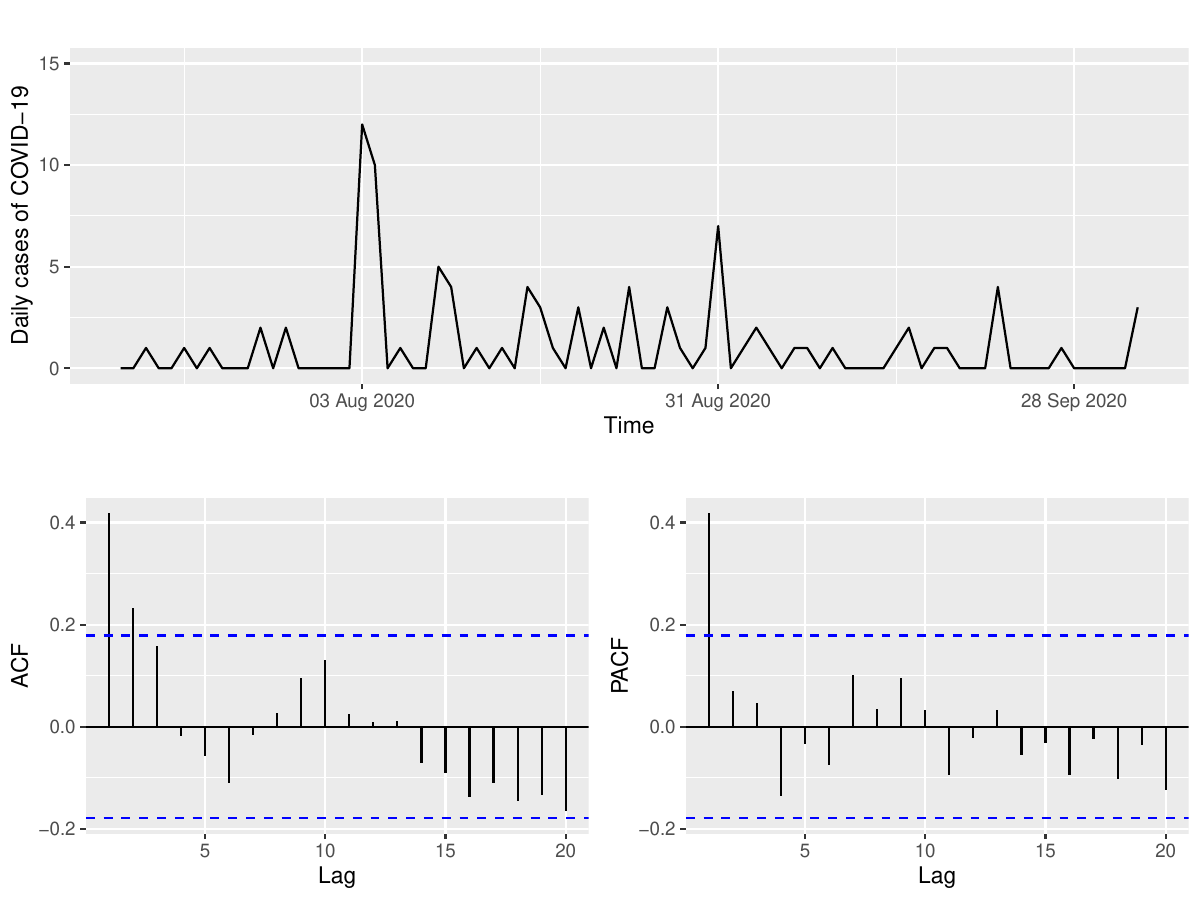}
		\caption{ Time series, ACF and PACF plots for Barbados COVID-19 data.}
		\label{fig:6}
	\end{figure}

\begin{figure}[H]
		\centering
		\includegraphics[scale=0.5]{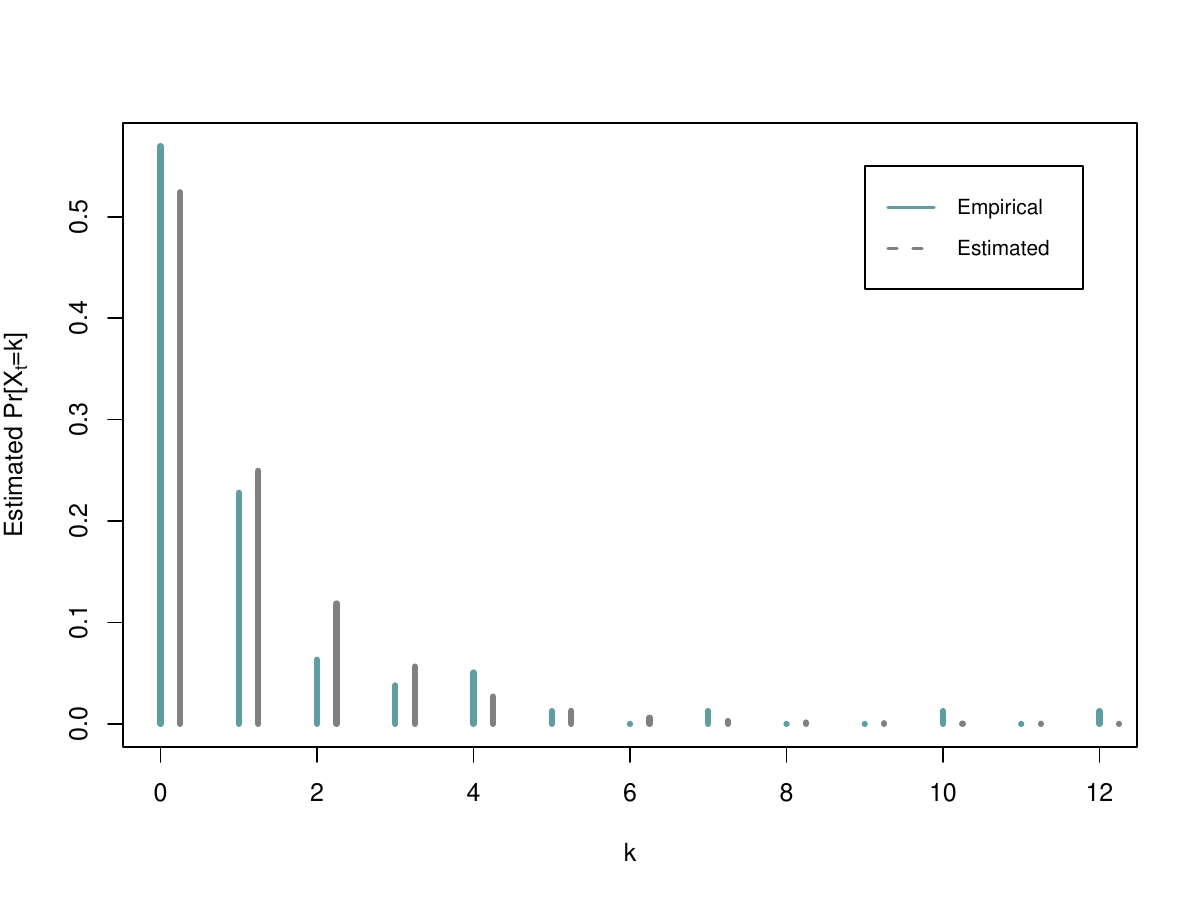}
		\caption{Plot of empirical frequencies( blue) and a geometric fit( gray) for Barbados COVID-19 data.}
		\label{fig:7}
	\end{figure}

	\noindent The first data comprises the daily counts of COVID-19 cases in Barbados, spanning from July 14, 2020, to September 30, 2020, encompassing 79 observations sourced from the World Health Organization (\href{https://covid19.who.int}{https://covid19.who.int}). The data exhibits overdispersion, as indicated by a mean of 1.1013 and a variance of 4.5024.  The autocorrelation function (ACF) and partial autocorrelation function (PACF) plots in \textcolor{blue}{Figure} \ref{fig:6} reveal an INAR(1) structure within the process. The marginal distribution of the data, as depicted in \textcolor{blue}{Figure} \ref{fig:7}, appears to exhibit skewness. Examination of Akaike information criterion (AIC) (\cite{aic}), Bayesian information criterion (BIC) (\cite{bic}) and corrected AIC (denoted by AIC$_c$) (\cite{aicc}) values in \textcolor{blue}{Table} \ref{tablas} indicate that the NoGeAR(1) model is the most suitable one for this data. In the present article, the last two data points have been considered as the `test set' and the first $77$ observations have been utilised for estimation and goodness of fit. 
	\begin{table}[H]
		\center
		\caption{Average PRMSE and PMAD when the data are from NoGeAR(1) model for h - step ahead forecasts }
		\renewcommand{\arraystretch}{1.6}
		\scalebox{0.66}{%
			\begin{tabular}{@{}lcccclcccc@{}}
				\toprule
				\multicolumn{10}{c}{\textbf{$n=140$, $m=60$}}                                                                                                                                                                          \\ \midrule
				\multicolumn{1}{c}{} & \multicolumn{4}{c}{\textbf{(I) $\alpha=0.8$, $\beta=0.2$, $\theta=0.5$}}      &                               & \multicolumn{4}{c}{\textbf{(III) $\alpha=0.6$, $\beta = 0.4$, $\theta = 0.75$}} \\ \midrule
				& \multicolumn{2}{c}{h=1}                & \multicolumn{2}{c}{h=2}              &                               & \multicolumn{2}{c}{h=1}                 & \multicolumn{2}{c}{h=2}               \\ \cmidrule(lr){2-5} \cmidrule(l){7-10} 
				\textbf{Model}       & PRMSE              & PMAD              & PRMSE             & PMAD             &                               & PRMSE              & PMAD               & PRMSE             & PMAD              \\ \cmidrule(r){1-5} \cmidrule(l){7-10} 
				NoGeAR(1)            & \textbf{2.6921}    & \textbf{1.4000}   & \textbf{2.7270}   & \textbf{1.4500}  &                               & \textbf{1.8086}    & \textbf{1.0833}    & \textbf{2.1677}   & \textbf{1.1167}   \\
				NGINAR               & 2.7294             & 1.7667            & 2.7712            & 1.8167           &                               & 1.8200             & 1.1196             & 2.1918            & 1.3917            \\
				GINAR                & 5.6035             & 2.0667            & 4.9207            & 2.4250           &                               & 2.9760             & 1.5917             & 3.1205            & 1.8667            \\
				PINAR                & 5.6580             & 3.9600            & 4.9400            & 4.1500           &                               & 5.2218             & 1.6083             & 5.2960            & 2.0667            \\ \cmidrule(r){1-5} \cmidrule(l){7-10} 
				\multicolumn{1}{c}{} & \multicolumn{4}{c}{\textbf{(II) $\alpha=0.7$, $\beta = 0.3$, $\theta = 0.5$}} &                               & \multicolumn{4}{c}{\textbf{(IV) $\alpha=0.55$. $\beta=0.45$, $\theta =0.83$}}   \\ \cmidrule(r){1-5} \cmidrule(l){7-10} 
				& \multicolumn{2}{c}{h=1}                & \multicolumn{2}{c}{h=2}              & \multicolumn{1}{c}{}          & \multicolumn{2}{c}{h=1}                 & \multicolumn{2}{c}{h=2}               \\ \cmidrule(lr){2-5} \cmidrule(l){7-10} 
				\textbf{Model}       & PRMSE              & PMAD              & PRMSE             & PMAD             &                               & PRMSE              & PMAD               & PRMSE             & PMAD              \\ \cmidrule(r){1-5} \cmidrule(l){7-10} 
				NoGeAR(1)            & \textbf{3.1122}    & \textbf{2.0917}   & \textbf{3.8768}   & \textbf{2.7583}  &                               & \textbf{1.0397}    & \textbf{0.4500}    & \textbf{1.2575}   & \textbf{0.7083}   \\
				NGINAR               & 3.1974             & 2.5333            & 3.9547            & 3.2167           &                               & 1.0404             & 0.7500             & 1.2629            & 0.7833            \\
				GINAR                & 8.7586             & 2.6250            & 8.8193            & 3.3333           &                               & 3.8869             & 0.6002             & 3.9429            & 0.8583            \\
				PINAR                & 6.5057             & 2.9745            & 6.5916            & 4.5833           &                               & 5.1715             & 0.8083             & 5.2493            & 0.9333            \\ \midrule
				\multicolumn{10}{c}{\textbf{$n=350$, $m=150$}}                                                                                                                                                                         \\ \midrule
				\multicolumn{1}{c}{} & \multicolumn{4}{c}{\textbf{(I) $\alpha=0.8$, $\beta=0.2$, $\theta =0.5$}}     &                               & \multicolumn{4}{l}{\textbf{(III) $\alpha=0.6$, $\beta = 0.4$, $\theta = 0.75$}} \\ \cmidrule(r){1-5} \cmidrule(l){7-10} 
				& \multicolumn{2}{c}{h=1}                & \multicolumn{2}{c}{h=2}              & \multicolumn{1}{c}{}          & \multicolumn{2}{c}{h=1}                 & \multicolumn{2}{c}{h=2}               \\ \cmidrule(lr){2-5} \cmidrule(l){7-10} 
				\textbf{Model}       & PRMSE              & PMAD              & PRMSE             & PMAD             &                               & PRMSE              & PMAD               & PRMSE             & PMAD              \\ \cmidrule(r){1-5} \cmidrule(l){7-10} 
				NoGeAR(1)            & \textbf{2.6579}    & \textbf{1.7525}   & \textbf{2.7111}            & \textbf{2.0025}  &                               & \textbf{1.8009}    & \textbf{1.2050}    & \textbf{2.1586}   & \textbf{1.4950}   \\
				NGINAR               & 2.6819             & 2.2125            & 2.7310            & 2.1550           &                               & 1.8118             & 1.2375             & 2.1915            & 1.6750            \\
				GINAR                & 4.8973             & 3.0175            & 4.9053            & 3.1225           &                               & 3.8763             & 1.4225             & 3.7128            & 1.8725            \\
				PINAR                & 4.9687             & 3.0575            & 4.9730            & 3.1075           &                               & 4.6225             & 1.7600             & 4.6375            & 1.9525            \\ \cmidrule(r){1-5} \cmidrule(l){7-10} 
				\multicolumn{1}{c}{} & \multicolumn{4}{c}{\textbf{(II) $\alpha=0.7$, $\beta = 0.3$, $\theta = 0.5$}} &                               & \multicolumn{4}{c}{\textbf{(IV) $\alpha=0.55$. $\beta=0.45$, $\theta =0.83$}}   \\ \cmidrule(r){1-5} \cmidrule(l){7-10} 
				& \multicolumn{2}{c}{h=1}                & \multicolumn{2}{c}{h=2}              & \multicolumn{1}{c}{\textbf{}} & \multicolumn{2}{c}{h=1}                 & \multicolumn{2}{c}{h=2}               \\ \cmidrule(lr){2-5} \cmidrule(l){7-10} 
				\textbf{Model}       & PRMSE              & PMAD              & PRMSE             & PMAD             & \textbf{}                     & PRMSE              & PMAD               & PRMSE             & PMAD              \\ \cmidrule(r){1-5} \cmidrule(l){7-10} 
				NoGeAR(1)            & \textbf{3.0390}    & \textbf{1.6800}   & \textbf{3.7543}   & \textbf{2.6075}  & \multicolumn{1}{c}{}          & \textbf{1.0328}    & \textbf{0.6275}    & \textbf{1.2496}   & \textbf{0.8375}   \\
				NGINAR               & 3.1353             & 1.9075            & 3.8664            & 2.6500           & \multicolumn{1}{c}{}          & 1.0613             & 0.6350             & 1.2775            & 0.8775            \\
				GINAR                & 8.8678             & 2.0150            & 8.8867            & 2.7275           & \multicolumn{1}{c}{}          & 4.8628             & 0.6625             & 4.867             & 0.9525            \\
				PINAR                & 6.7845             & 2.9950            & 6.8048            & 4.0525           & \multicolumn{1}{c}{}          & 6.8177             & 1.0250             & 6.8796            & 1.0450            \\ \bottomrule
		\end{tabular}}
		\label{tab2}
	\end{table}
	
	\begin{table}[H]
		\center
		\caption{PTP comparison of NoGeAR(1) and NGINAR models data are from NoGeAR(1)}
		\renewcommand{\arraystretch}{2}
		\scalebox{0.7}{%
			\begin{tabular}{lcccclcccc}
				\hline
				\multicolumn{10}{c}{\textbf{$n=140$, $m=60$}}                                                                                                                                                                                                                                                   \\ \hline
				\multicolumn{1}{c}{} & \multicolumn{4}{c}{\textbf{(I) $\alpha=0.8$, $\beta=0.2$, $\theta =0.5$}}                                               &                      & \multicolumn{4}{c}{\textbf{(III) $\alpha=0.6$, $\beta = 0.4$, $\theta = 0.75$}}                                         \\ \hline
				& \multicolumn{2}{c}{h=1}                                    & \multicolumn{2}{c}{h=2}                                    &                      & \multicolumn{2}{c}{h=1}                                    & \multicolumn{2}{c}{h=2}                                    \\ \cline{2-5} \cline{7-10} 
				& \multicolumn{1}{l}{NoGeAR(1)} & \multicolumn{1}{l}{NGINAR} & \multicolumn{1}{l}{NoGeAR(1)} & \multicolumn{1}{l}{NGINAR} &                      & \multicolumn{1}{l}{NoGeAR(1)} & \multicolumn{1}{l}{NGINAR} & \multicolumn{1}{l}{NoGeAR(1)} & \multicolumn{1}{l}{NGINAR} \\ \cline{1-5} \cline{7-10} 
				Mean                 & 23.68                         & 24.93                      & 24.40                         & 25.38                      & \multicolumn{1}{c}{} & 18.28                         & 17.42                      & 13.32                         & 13.78                      \\
				Median               & 35.83                         & 49.17                      & 43.33                         & 50.00                      & \multicolumn{1}{c}{} & 32.50                         & 23.33                      & 25.00                         & 11.67                      \\
				Mode                 & 44.17                         & 56.67                      & 40.00                         & 55.00                      & \multicolumn{1}{c}{} & 38.33                         & 28.33                      & 25.83                         & 25.00                      \\ \cline{1-5} \cline{7-10} 
				\multicolumn{1}{c}{} & \multicolumn{4}{c}{\textbf{(II) $\alpha=0.7$, $\beta = 0.3$, $\theta = 0.5$}}                                           &                      & \multicolumn{4}{c}{\textbf{(IV) $\alpha=0.55$. $\beta=0.45$, $\theta =0.83$}}                                           \\ \cline{1-5} \cline{7-10} 
				& \multicolumn{2}{c}{h=1}                                    & \multicolumn{2}{c}{h=2}                                    & \multicolumn{1}{c}{} & \multicolumn{2}{c}{h=1}                                    & \multicolumn{2}{c}{h=2}                                    \\ \cline{2-5} \cline{7-10} 
				& \multicolumn{1}{l}{NoGeAR(1)} & \multicolumn{1}{l}{NGINAR} & \multicolumn{1}{l}{NoGeAR(1)} & \multicolumn{1}{l}{NGINAR} &                      & \multicolumn{1}{l}{NoGeAR(1)} & \multicolumn{1}{l}{NGINAR} & \multicolumn{1}{l}{NoGeAR(1)} & \multicolumn{1}{l}{NGINAR} \\ \cline{1-5} \cline{7-10} 
				Mean                 & 28.08                         & 27.17                      & 24.07                         & 24.68                      &                      & 14.55                         & 16.10                      & 10.76                         & 11.35                      \\
				Median               & 48.33                         & 40.83                      & 49.17                         & 42.50                      &                      & 24.50                         & 25.83                      & 14.17                         & 10.00                      \\
				Mode                 & 50.00                         & 43.33                      & 49.17                         & 47.50                      &                      & 24.17                         & 25.00                      & 13.33                         & 17.50                      \\ \hline
				\multicolumn{10}{c}{\textbf{$n=350$, $m=150$}}                                                                                                                                                                                                                                                  \\ \hline
				\multicolumn{1}{c}{} & \multicolumn{4}{c}{\textbf{(I) $\alpha=0.8$, $\beta=0.2$, $\theta =0.5$}}                                               &                      & \multicolumn{4}{c}{\textbf{(III) $\alpha=0.6$, $\beta = 0.4$, $\theta = 0.75$}}                                         \\ \cline{1-5} \cline{7-10} 
				& \multicolumn{2}{c}{h=1}                                    & \multicolumn{2}{c}{h=2}                                    & \multicolumn{1}{c}{} & \multicolumn{2}{c}{h=1}                                    & \multicolumn{2}{c}{h=2}                                    \\ \cline{2-5} \cline{7-10} 
				& \multicolumn{1}{l}{NoGeAR(1)} & \multicolumn{1}{l}{NGINAR} & \multicolumn{1}{l}{NoGeAR(1)} & \multicolumn{1}{l}{NGINAR} &                      & \multicolumn{1}{l}{NoGeAR(1)} & \multicolumn{1}{l}{NGINAR} & \multicolumn{1}{l}{NoGeAR(1)} & \multicolumn{1}{l}{NGINAR} \\ \cline{1-5} \cline{7-10} 
				Mean                 & 24.74                         & 24.00                      & 25.51                         & 24.69                      &                      & 17.47                         & 17.12                      & 12.88                         & 12.62                      \\
				Median               & 44.00                         & 39.00                      & 45.33                         & 44.00                      &                      & 28.67                         & 17.00                      & 18.00                         & 15.67                      \\
				Mode                 & 50.33                         & 48.00                      & 33.67                         & 48.67                      &                      & 28.67                         & 21.00                      & 17.67                         & 25.00                      \\ \cline{1-5} \cline{7-10} 
				\multicolumn{1}{c}{} & \multicolumn{4}{c}{\textbf{(II) $\alpha=0.7$, $\beta = 0.3$, $\theta = 0.5$}}                                           &                      & \multicolumn{4}{c}{\textbf{(IV) $\alpha=0.55$. $\beta=0.45$, $\theta =0.83$}}                                           \\ \cline{1-5} \cline{7-10} 
				& \multicolumn{2}{c}{h=1}                                    & \multicolumn{2}{c}{h=2}                                    & \multicolumn{1}{c}{} & \multicolumn{2}{c}{h=1}                                    & \multicolumn{2}{c}{h=2}                                    \\ \cline{2-5} \cline{7-10} 
				& \multicolumn{1}{l}{NoGeAR(1)} & \multicolumn{1}{l}{NGINAR} & \multicolumn{1}{l}{NoGeAR(1)} & \multicolumn{1}{l}{NGINAR} &                      & \multicolumn{1}{l}{NoGeAR(1)} & \multicolumn{1}{l}{NGINAR} & \multicolumn{1}{l}{NoGeAR(1)} & \multicolumn{1}{l}{NGINAR} \\ \cline{1-5} \cline{7-10} 
				Mean                 & 25.83                         & 26.87                      & 24.57                         & 24.71                      & \multicolumn{1}{c}{} & 15.66                         & 15.39                      & 11.30                         & 11.05                      \\
				Median               & 50.67                         & 34.67                      & 49.33                         & 36.67                      & \multicolumn{1}{c}{} & 19.67                         & 18.67                      & 14.67                         & 19.00                      \\
				Mode                 & 52.67                         & 34.33                      & 50.66                         & 43.67                      & \multicolumn{1}{c}{} & 20.00                         & 20.33                      & 14.67                         & 22.33                      \\ \hline
		\end{tabular}}
		\label{tab3}
	\end{table}

	\begin{table}[H]
		\center
		\caption{Average PRMSE and PMAD when the data are from NGINAR model for h - step ahead forecasts}
		\renewcommand{\arraystretch}{1.5}
		\scalebox{0.8}{%
			\begin{tabular}{lcccclcccc}
				\hline
				\multicolumn{10}{c}{\textbf{$n=140$, $m=60$}}                                                                                                                                                        \\ \hline
				\multicolumn{1}{c}{} & \multicolumn{4}{c}{\textbf{(I) $\alpha^{ng}=0.67$, $\mu=3$}}               &                               & \multicolumn{4}{c}{\textbf{(III) $\alpha^{ng}=0.43$, $\mu =1 $}}           \\ \hline
				& \multicolumn{2}{c}{h=1}           & \multicolumn{2}{c}{h=2}           &                               & \multicolumn{2}{c}{h=1}           & \multicolumn{2}{c}{h=2}           \\ \cline{2-5} \cline{7-10} 
				\textbf{Model}       & PRMSE           & PMAD            & PRMSE           & PMAD            &                               & PRMSE           & PMAD            & PRMSE           & PMAD            \\ \cline{1-5} \cline{7-10} 
				NoGeAR(1)            & \textbf{2.5644} & \textbf{1.5250} & \textbf{3.1038} & 2.2917          &                               & 1.2720          & 0.7833          & 1.3701          & \textbf{0.7750} \\
				NGINAR               & 2.6053          & 1.6333          & 3.1695          & \textbf{2.1000} &                               & \textbf{1.2476} & 1.1500          & \textbf{1.3528} & 1.0250          \\
				GINAR                & 3.4831          & 1.9000          & 3.5042          & 2.3250          &                               & 1.4165          & \textbf{0.7250} & 1.4206          & 0.7833          \\
				PINAR                & 5.5082          & 2.6167          & 5.5763          & 3.9333          &                               & 6.7273          & 0.8083          & 6.8023          & 0.8250          \\ \cline{1-5} \cline{7-10} 
				\multicolumn{1}{c}{} & \multicolumn{4}{c}{\textbf{(II) $\alpha^{ng}=0.25$, $\mu =1$}}             &                               & \multicolumn{4}{c}{\textbf{(IV) $\alpha^{ng}=0.82$. $\mu=5$}}              \\ \cline{1-5} \cline{7-10} 
				& \multicolumn{2}{c}{h=1}           & \multicolumn{2}{c}{h=2}           & \multicolumn{1}{c}{}          & \multicolumn{2}{c}{h=1}           & \multicolumn{2}{c}{h=2}           \\ \cline{2-5} \cline{7-10} 
				\textbf{Model}       & PRMSE           & PMAD            & PRMSE           & PMAD            &                               & PRMSE           & PMAD            & PRMSE           & PMAD            \\ \cline{1-5} \cline{7-10} 
				NoGeAR(1)            & 1.3831          & \textbf{0.7833} & 1.4142          & 0.9917          &                               & 3.0983          & \textbf{1.6000} & 4.0077          & \textbf{2.3750} \\
				NGINAR               & \textbf{1.3405} & 0.9917          & \textbf{1.3764} & \textbf{0.9750} &                               & \textbf{3.0529} & 3.6333          & \textbf{3.9954} & 5.4667          \\
				GINAR                & 1.3757          & 0.9583          & 1.3766          & 1.0750          &                               & 5.5406          & 2.3667          & 5.5875          & 2.9917          \\
				PINAR                & 9.5672          & 1.0833          & 9.6778          & 1.0750          &                               & 7.7173          & 2.4000          & 7.8783          & 2.8833          \\ \hline
				\multicolumn{10}{c}{\textbf{$n=350$, $m=150$}}                                                                                                                                                       \\ \hline
				\multicolumn{1}{c}{} & \multicolumn{4}{c}{\textbf{(I) $\alpha^{ng}=0.67$, $\mu=3$}}               &                               & \multicolumn{4}{c}{\textbf{(III) $\alpha^{ng}=0.43$, $\mu =1 $}}           \\ \cline{1-5} \cline{7-10} 
				& \multicolumn{2}{c}{h=1}           & \multicolumn{2}{c}{h=2}           & \multicolumn{1}{c}{}          & \multicolumn{2}{c}{h=1}           & \multicolumn{2}{c}{h=2}           \\ \cline{2-5} \cline{7-10} 
				\textbf{Model}       & PRMSE           & PMAD            & PRMSE           & PMAD            &                               & PRMSE           & PMAD            & PRMSE           & PMAD            \\ \cline{1-5} \cline{7-10} 
				NoGeAR(1)            & \textbf{2.4746} & \textbf{1.7233} & \textbf{2.9936} & 2.7500          &                               & 1.2954          & 0.9000          & 1.4074          & 1.0367          \\
				NGINAR               & 2.5503          & 2.4733          & 3.0781          & 2.6500          &                               & \textbf{1.2689} & 1.1733          & \textbf{1.3797} & 0.9367          \\
				GINAR                & 3.4581          & 2.0100          & 3.4644          & 2.4367          &                               & 1.3959          & \textbf{0.7167} & 1.3967          & \textbf{0.8467} \\
				PINAR                & 5.5497          & 1.8133          & 5.5676          & \textbf{2.1800} &                               & 8.0241          & 0.8433          & 8.0452          & 0.9833          \\ \cline{1-5} \cline{7-10} 
				\multicolumn{1}{c}{} & \multicolumn{4}{c}{\textbf{(II) $\alpha^{ng}=0.25$, $\mu =1$}}             &                               & \multicolumn{4}{c}{\textbf{(IV) $\alpha^{ng}=0.82$. $\mu=5$}}              \\ \cline{1-5} \cline{7-10} 
				& \multicolumn{2}{c}{h=1}           & \multicolumn{2}{c}{h=2}           & \multicolumn{1}{c}{\textbf{}} & \multicolumn{2}{c}{h=1}           & \multicolumn{2}{c}{h=2}           \\ \cline{2-5} \cline{7-10} 
				\textbf{Model}       & PRMSE           & PMAD            & PRMSE           & PMAD            & \textbf{}                     & PRMSE           & PMAD            & PRMSE           & PMAD            \\ \cline{1-5} \cline{7-10} 
				NoGeAR(1)            & \textbf{1.3641} & \textbf{0.7500} & \textbf{1.4096} & \textbf{0.8533} & \multicolumn{1}{c}{}          & 3.1351          & 2.5100              & 4.0947          & 4.9533          \\
				NGINAR               & 1.3686          & 0.9900          & 1.4142          & 0.9867          & \multicolumn{1}{c}{}          & \textbf{3.1337} & 3.7367          & \textbf{4.0713} & 5.2767          \\
				GINAR                & 1.4228          & 0.8300          & 1.4234          & 0.9767          & \multicolumn{1}{c}{}          & 5.3247          & \textbf{2.2867} & 5.3349          & \textbf{3.0200} \\
				PINAR                & 2.6145          & 1.0200          & 2.6178          & 1.0467          & \multicolumn{1}{c}{}          & 8.3084          & 3.4000          & 8.3310          & 4.0700          \\ \hline
			\end{tabular}
		}
		\label{tab4}
	\end{table}
	
	\begin{table}[H]
		\center
		\caption{PTP comparison of NoGeAR(1) and NGINAR models data are from NGINAR }
		\renewcommand{\arraystretch}{2}
		\scalebox{0.7}{%
			\begin{tabular}{lcccclcccc}
				\hline
				\multicolumn{10}{c}{\textbf{$n=140$, $m=60$}}                                                                                                                                                                                                                                                   \\ \hline
				\multicolumn{1}{c}{} & \multicolumn{4}{c}{\textbf{(I) $\alpha^{ng}=0.67$, $\mu=3$}}                                                                 &                      & \multicolumn{4}{c}{\textbf{(III) $\alpha^{ng}=0.43$, $\mu =1 $}}                                                             \\ \hline
				& \multicolumn{2}{c}{h=1}                                    & \multicolumn{2}{c}{h=2}                                    &                      & \multicolumn{2}{c}{h=1}                                    & \multicolumn{2}{c}{h=2}                                    \\ \cline{2-5} \cline{7-10} 
				& \multicolumn{1}{l}{NoGeAR(1)} & \multicolumn{1}{l}{NGINAR} & \multicolumn{1}{l}{NoGeAR(1)} & \multicolumn{1}{l}{NGINAR} &                      & \multicolumn{1}{l}{NoGeAR(1)} & \multicolumn{1}{l}{NGINAR} & \multicolumn{1}{l}{NoGeAR(1)} & \multicolumn{1}{l}{NGINAR} \\ \cline{1-5} \cline{7-10} 
				Mean                 & 18.00                         & 17.56                      & 13.92                         & 12.63                      & \multicolumn{1}{c}{} & 29.85                         & 29.68                      & 24.83                         & 25.55                      \\
				Median               & 33.33                         & 31.67                      & 25.00                         & 20.00                      & \multicolumn{1}{c}{} & 43.33                         & 34.17                      & 55.00                         & 49.17                      \\
				Mode                 & 35.00                         & 32.50                      & 38.33                         & 29.17                      & \multicolumn{1}{c}{} & 50.83                         & 43.33                      & 53.33                         & 43.33                      \\ \cline{1-5} \cline{7-10} 
				\multicolumn{1}{c}{} & \multicolumn{4}{c}{\textbf{(II) $\alpha^{ng}=0.25$, $\mu =1$}}                                                               &                      & \multicolumn{4}{c}{\textbf{(IV) $\alpha^{ng}=0.82$. $\mu=5$}}                                                                \\ \cline{1-5} \cline{7-10} 
				& \multicolumn{2}{c}{h=1}                                    & \multicolumn{2}{c}{h=2}                                    & \multicolumn{1}{c}{} & \multicolumn{2}{c}{h=1}                                    & \multicolumn{2}{c}{h=2}                                    \\ \cline{2-5} \cline{7-10} 
				& \multicolumn{1}{l}{NoGeAR(1)} & \multicolumn{1}{l}{NGINAR} & \multicolumn{1}{l}{NoGeAR(1)} & \multicolumn{1}{l}{NGINAR} &                      & \multicolumn{1}{l}{NoGeAR(1)} & \multicolumn{1}{l}{NGINAR} & \multicolumn{1}{l}{NoGeAR(1)} & \multicolumn{1}{l}{NGINAR} \\ \cline{1-5} \cline{7-10} 
				Mean                 & 23.47                         & 24.62                      & 24.33                         & 25.25                      &                      & 15.53                         & 14.58                      & 10.75                         & 10.62                      \\
				Median               & 56.67                         & 40.83                      & 41.67                         & 42.50                      &                      & 20.83                         & 21.67                      & 18.33                         & 11.67                      \\
				Mode                 & 48.33                         & 49.17                      & 40.00                         & 47.50                      &                      & 31.67                         & 25.83                      & 18.33                         & 24.17                      \\ \hline
				\multicolumn{10}{c}{\textbf{$n=350$, $m=150$}}                                                                                                                                                                                                                                                  \\ \hline
				\multicolumn{1}{c}{} & \multicolumn{4}{c}{\textbf{(I) $\alpha^{ng}=0.67$, $\mu=3$}}                                                                 &                      & \multicolumn{4}{c}{\textbf{(III) $\alpha^{ng}=0.43$, $\mu =1 $}}                                                             \\ \cline{1-5} \cline{7-10} 
				& \multicolumn{2}{c}{h=1}                                    & \multicolumn{2}{c}{h=2}                                    & \multicolumn{1}{c}{} & \multicolumn{2}{c}{h=1}                                    & \multicolumn{2}{c}{h=2}                                    \\ \cline{2-5} \cline{7-10} 
				& \multicolumn{1}{l}{NoGeAR(1)} & \multicolumn{1}{l}{NGINAR} & \multicolumn{1}{l}{NoGeAR(1)} & \multicolumn{1}{l}{NGINAR} &                      & \multicolumn{1}{l}{NoGeAR(1)} & \multicolumn{1}{l}{NGINAR} & \multicolumn{1}{l}{NoGeAR(1)} & \multicolumn{1}{l}{NGINAR} \\ \cline{1-5} \cline{7-10} 
				Mean                 & 17.45                         & 17.63                      & 12.98                         & 13.17                      &                      & 25.67                         & 27.04                      & 24.22                         & 24.77                      \\
				Median               & 26.67                         & 22.00                      & 15.67                         & 13.33                      &                      & 43.67                         & 42.67                      & 46.00                         & 41.67                      \\
				Mode                 & 24.33                         & 24.00                      & 19.00                         & 25.67                      &                      & 51.00                         & 40.00                      & 44.33                         & 56.00                      \\ \cline{1-5} \cline{7-10} 
				\multicolumn{1}{c}{} & \multicolumn{4}{c}{\textbf{(II) $\alpha^{ng}=0.25$, $\mu =1$}}                                                               &                      & \multicolumn{4}{c}{\textbf{(IV) $\alpha^{ng}=0.82$. $\mu=5$}}                                                                \\ \cline{1-5} \cline{7-10} 
				& \multicolumn{2}{c}{h=1}                                    & \multicolumn{2}{c}{h=2}                                    & \multicolumn{1}{c}{} & \multicolumn{2}{c}{h=1}                                    & \multicolumn{2}{c}{h=2}                                    \\ \cline{2-5} \cline{7-10} 
				& \multicolumn{1}{l}{NoGeAR(1)} & \multicolumn{1}{l}{NGINAR} & \multicolumn{1}{l}{NoGeAR(1)} & \multicolumn{1}{l}{NGINAR} &                      & \multicolumn{1}{l}{NoGeAR(1)} & \multicolumn{1}{l}{NGINAR} & \multicolumn{1}{l}{NoGeAR(1)} & \multicolumn{1}{l}{NGINAR} \\ \cline{1-5} \cline{7-10} 
				Mean                 & 24.81                         & 23.97                      & 25.018                        & 25.022                     & \multicolumn{1}{c}{} & 15.39                         & 15.25                      & 11.24                         & 11.17                      \\
				Median               & 49.67                         & 38.67                      & 51.33                         & 43.33                      & \multicolumn{1}{c}{} & 24.33                            & 14.33                      & 12.33                         & 10.00                      \\
				Mode                 & 53.33                         & 46.67                      & 53.00                         & 48.00                      & \multicolumn{1}{c}{} & 22.67                         & 23.67                      & 14.00                         & 18.67                      \\ \hline
			\end{tabular}
		}
		\label{tab5}
	\end{table}

		\begin{table}[H]
		\center
		\caption{P-values of Portmanteau test statistics for Barbados cases for various values of m}
		\renewcommand{\arraystretch}{1}
		\scalebox{0.8}{%
			\begin{tabular}{@{}llllll@{}}
				\toprule
				{m}        & {2}    & {3}    & {4}    & {8}    & {12}   \\ \midrule
				{ $Q_{BP}$} & {0.38} & {0.56} & {0.31} & { 0.27} & {0.48} \\
				{$Q_{LB}$} & {0.36} & {0.54} & {0.29} & {0.20} & {0.38} \\
				{$Q_{LM}$} & {0.37} & {0.55} & {0.29} & {0.23} & {0.40} \\ \bottomrule
		\end{tabular}}
		\label{port}
	\end{table}
	
\begin{figure}[H]
		\centering
		\includegraphics[scale=0.7]{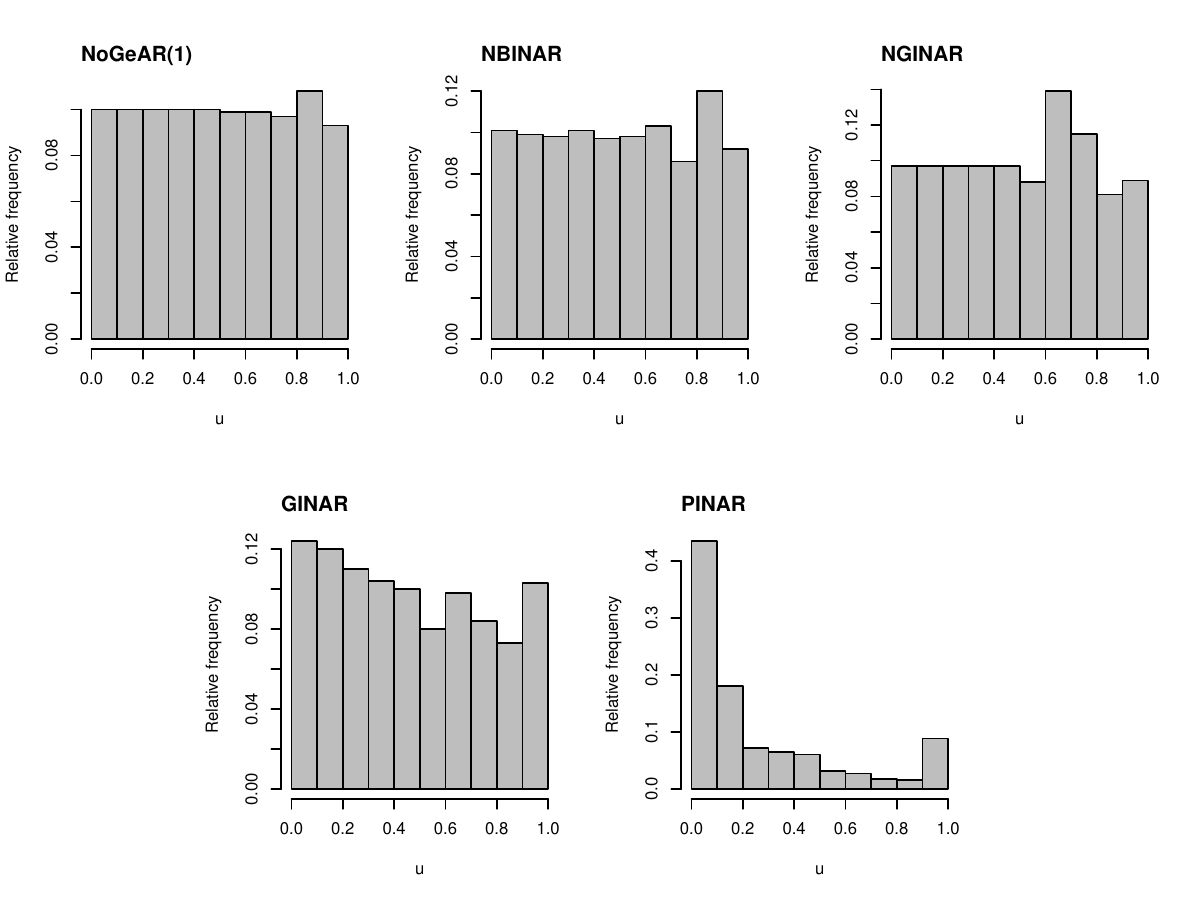}
		\caption{ PIT histograms following analyis of Barbados COVID-19 data.}
		\label{fig:8}
	\end{figure}

\begin{figure}[t]
		\centering
		\includegraphics[scale=0.6]{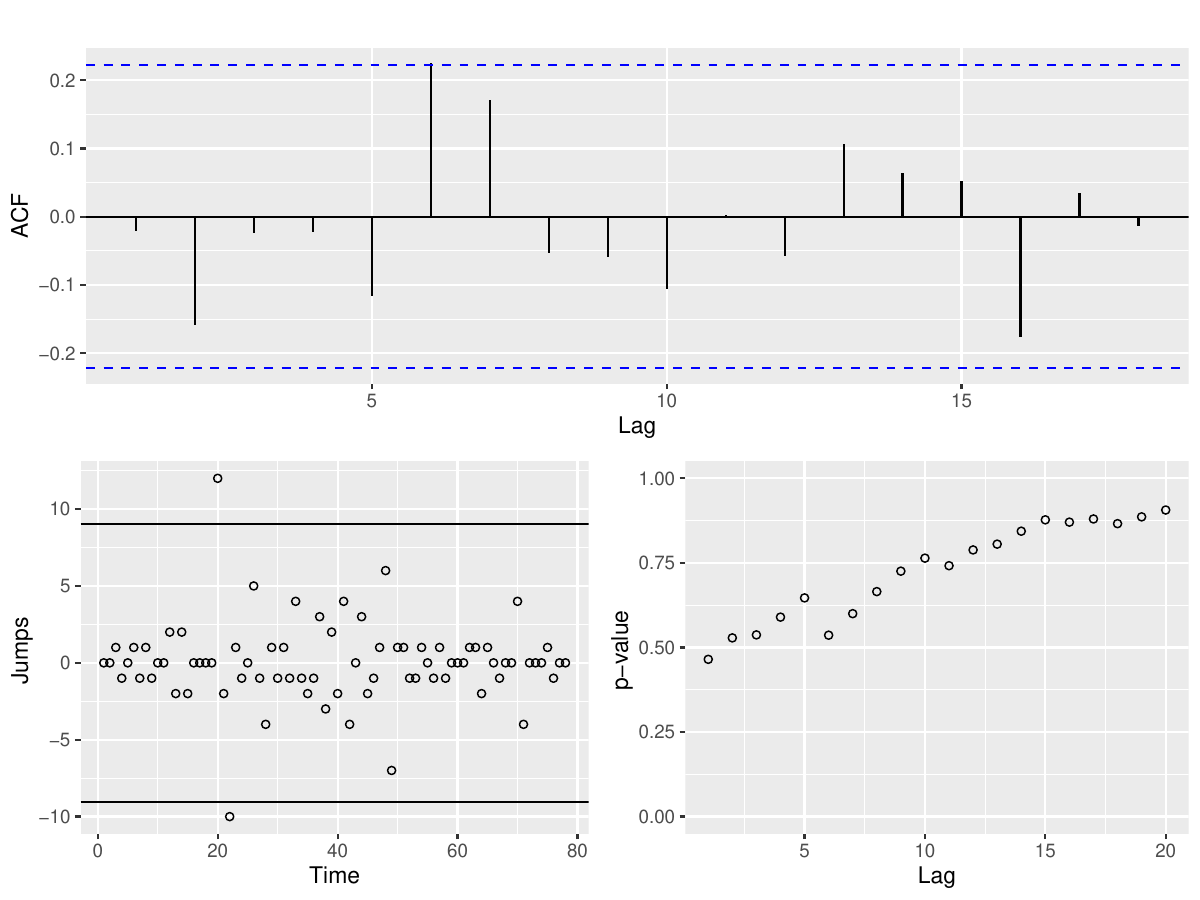}
		\caption{  Plots of ACF of the residuals, Jumps against time, and Ljung–Box p- values for the Barbados COVID-19 data.}
		\label{fig:9}
	\end{figure}

	In \textcolor{blue}{Table} \ref{last} the point forecasts for the last two observations, i.e. the one-step ahead forecast of the 77th observation and the two-step ahead forecast of the 76th observation, along with the corresponding HPP intervals with 0.95 coverage probability are provided for various INAR(1) models. Though NoGeAR(1) model proved to be the best fit among the models compared, the median point forecasts of NBINAR are shown to be equal to the actual values. This could be due to the capturing of noise along with the observations by NoGeAR(1) model. The probability integral transform (PIT) histogram (See \cite{weiss2018introduction}) in \textcolor{blue}{Figure} \ref{fig:8} displays almost uniform distribution and the lack of dependence of Pearson's residuals reflected in the ACF plot  in \textcolor{blue}{Figure} \ref{fig:9} support the appropriateness of the model fit for the data. Moreover, the jumps against time control chart in \textcolor{blue}{Figure} \ref{fig:9}, where jumps are given by $X_t -X_{t-1}$, imply model adequacy as most of the observations lie within the control limits $\pm 3\sigma_J$ ($\sigma_J$ being the standard deviation of jumps) (See \cite{li2016effective}, \cite{guerrero2022integer}).  
	\subsection{\TeX\ editor downloads}
\begin{figure}[H]
		\centering
		\includegraphics[scale=0.5]{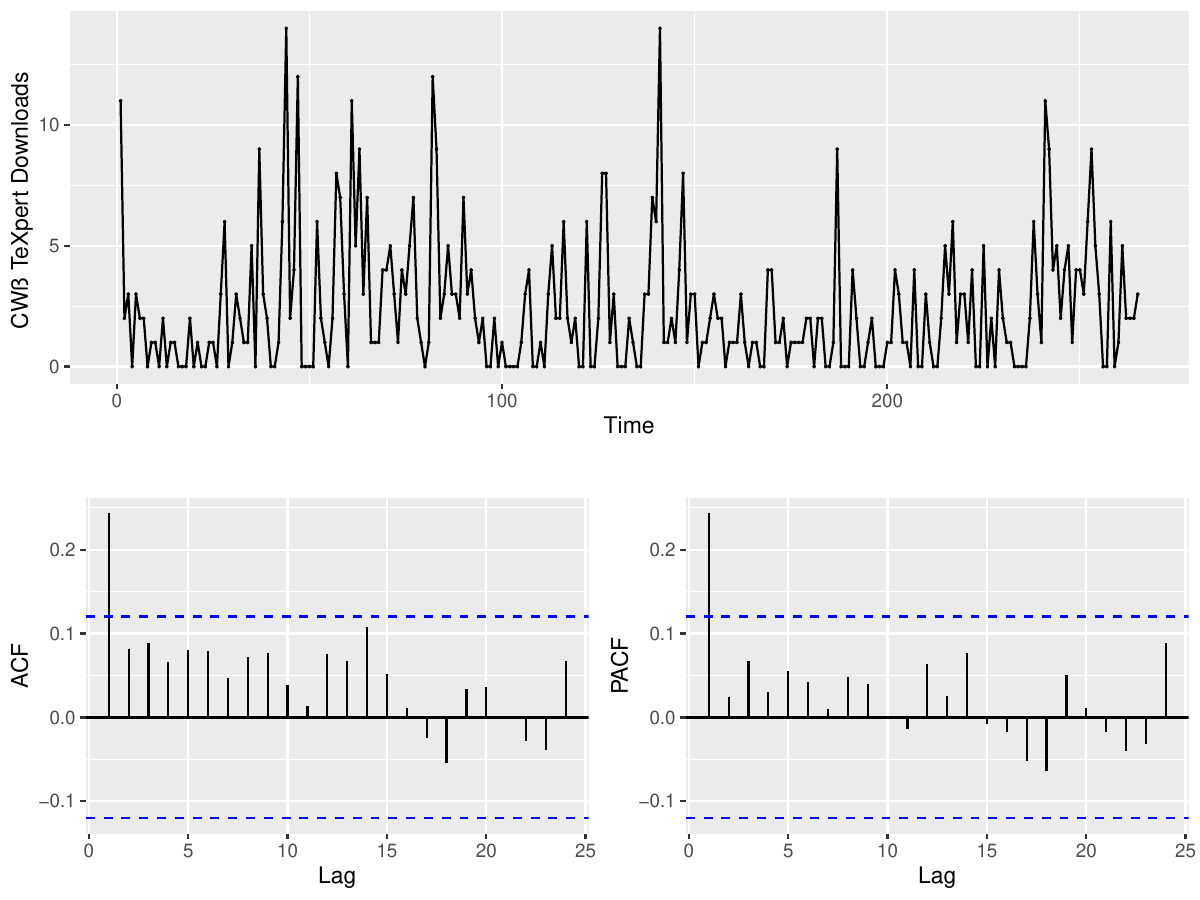}
		\caption{ Time series, ACF and PACF plots for CWß TeXpert downloads data}
		\label{fig:2}
	\end{figure}

	The software CWß TeXpert, a no-cost \TeX\ editor designed for Windows, has been accessible since 2003 via \href{https://www.berlios.de}{www.berlios.de} (See \cite{weiss2008thinning}). The program's download statistics are recorded each day when the server is operational. \cite{doi:10.1080/00949655.2023.2213794} analysed the data recorded from June 1, 2006, to February 28, 2007 and concluded that NoGeAR(1) model fits the data well. The PACF plot in \textcolor{blue}{Figure} \ref{fig:2} provides a basis for validating the AR(1) structure assumptions of the model. The results are in line with those presented by \cite{doi:10.1080/00949655.2023.2213794} and is clear from the AIC, BIC and AIC$_c$ values for various models in \textcolor{blue}{Table} \ref{tab6}.\par
	
\begin{figure}[H]
		\centering
		\includegraphics[scale=0.6]{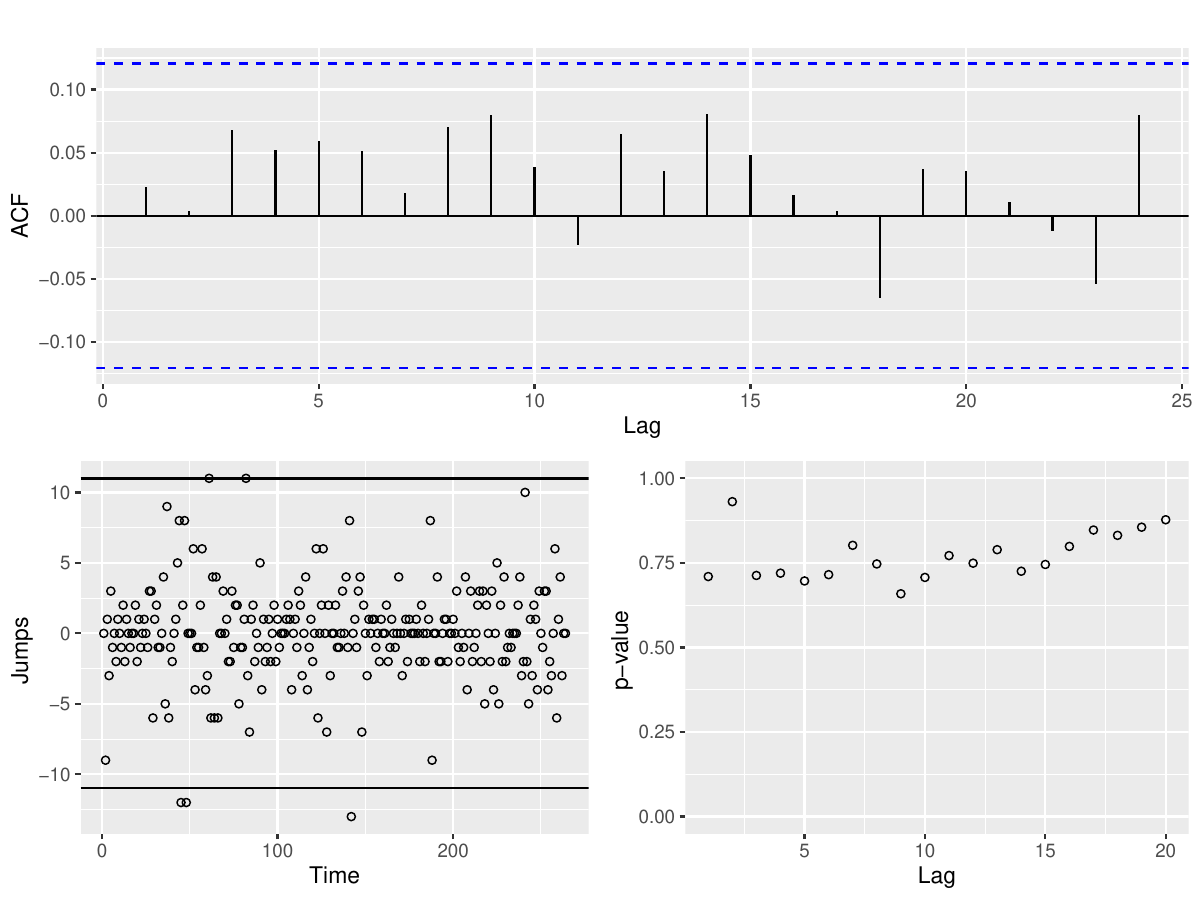}
		\caption{  Plots of ACF of the residuals, Jumps against time, and Ljung–Box p- values for the Downloads dataset.}
		\label{fig:5}
	\end{figure}

\begin{figure}[H]
		\centering
		\includegraphics[scale=0.6]{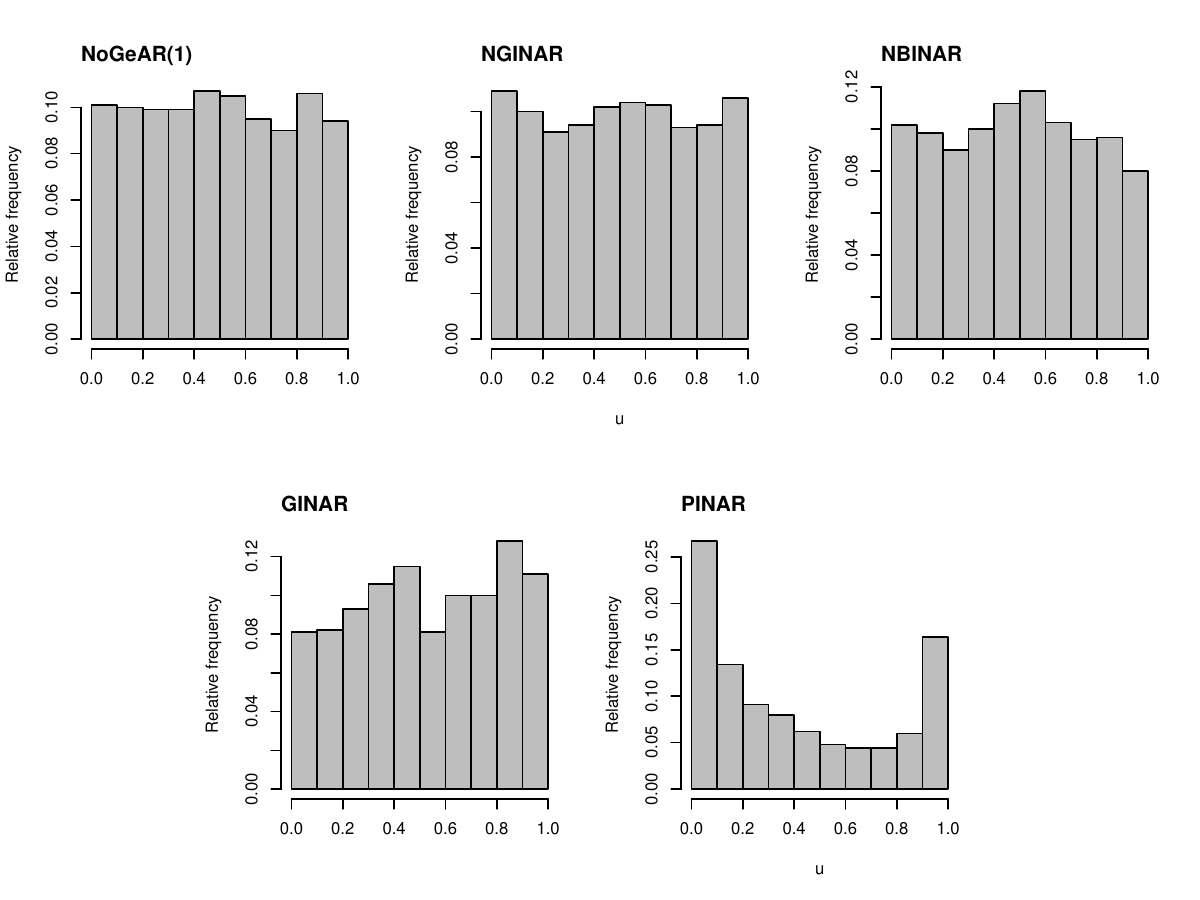}
		\caption{ PIT histograms following analyis of  CWß TeXpert downloads data.}
		\label{fig:4}
	\end{figure}

	\begin{table}[H]
		\center
		\caption{Saddlepoint Approximation Estimates, AIC, BIC and  AIC$_c$ values for various models fitted to Barbados COVID-19 data }
		\renewcommand{\arraystretch}{1.5}
		\scalebox{0.68}{%
			\begin{tabular}{@{}llccccl@{}}
				\toprule
				\multirow{2}{*}{\textbf{Model}} & \multicolumn{3}{c}{\textbf{Parameter}}                                                                              & \multirow{2}{*}{\textbf{AIC}} & \multicolumn{1}{l}{\multirow{2}{*}{\textbf{BIC}}} & \multirow{2}{*}{\textbf{AIC$_c$}} \\ \cmidrule(lr){2-4}
				& 1                                           & 2                                           & 3                       &                               & \multicolumn{1}{l}{}                              &                                   \\ \midrule
				NoGeAR(1)                       & \multirow{2}{*}{0.9745}                     & \multirow{2}{*}{0.5260}                     & \multirow{2}{*}{0.5543} & \multirow{2}{*}{204.69}       & \multirow{2}{*}{207.06}                           & \multirow{2}{*}{204.85}           \\
				($\alpha, \beta, \theta$)       &                                             &                                             &                         &                               &                                                   &                                   \\
				NGINAR                          & \multirow{2}{*}{1.1744}                     & \multirow{2}{*}{0.1403}                     & \multirow{2}{*}{}       & \multirow{2}{*}{232.71}       & \multirow{2}{*}{239.82}                           & \multirow{2}{*}{233.03}           \\
				($\mu, \alpha$)                 &                                             &                                             &                         &                               &                                                   &                                   \\
				NBINAR                          & \multirow{2}{*}{0.4111}                     & \multirow{2}{*}{0.1277}                     & \multirow{2}{*}{0.7300} & \multirow{2}{*}{231.81}       & \multirow{2}{*}{236.55}                           & \multirow{2}{*}{231.97}           \\
				($n, p, \alpha$)                &                                             &                                             &                         &                               &                                                   &                                   \\
				GINAR                           & \multicolumn{1}{c}{\multirow{2}{*}{0.5214}} & \multirow{2}{*}{0.0589}                     & \multirow{2}{*}{}       & \multirow{2}{*}{235.82}       & \multirow{2}{*}{237.19}                           & \multirow{2}{*}{236.14}           \\
				($p, \alpha$)                   & \multicolumn{1}{c}{}                        &                                             &                         &                               &                                                   &                                   \\
				PINAR                           & \multirow{2}{*}{1.1154}                     & \multicolumn{1}{l}{\multirow{2}{*}{0.0589}} & \multirow{2}{*}{}       & \multirow{2}{*}{294.37}       & \multirow{2}{*}{299.11}                           & \multirow{2}{*}{294.53}           \\
				($\lambda, \alpha$)             &                                             & \multicolumn{1}{l}{}                        &                         &                               &                                                   &                                   \\ \bottomrule
			\end{tabular}
		}
		\label{tablas}
	\end{table}
		The uncorrelatedness of Pearson's residuals is affirmed by the ACF plot in \textcolor{blue}{Figure} \ref{fig:5}, and the PIT histogram in \textcolor{blue}{Figure} \ref{fig:4} shows near uniformity, supporting the adequacy of the model for the data. \textcolor{blue}{Table} \ref{tab7} presents point forecasts and HPP intervals for the last two observations. It is observed that the median forecasts of NoGeAR(1) model provide better forecasts as compared to the other models.\par
	
	\begin{table}[H]
		\center
		\caption{Point forecasts, HPP intervals of $95\%$ coverage probability for Barbados COVID-19 data}
		\renewcommand{\arraystretch}{1.5}
		\scalebox{0.8}{%
	\begin{tabular}{lcccccc}
		\hline
		& \multicolumn{1}{l}{}                                  &                                                                                    & \multicolumn{1}{l}{}                                & \multicolumn{1}{l}{}                                  & \multicolumn{1}{l}{}                                & \multicolumn{1}{l}{{\color[HTML]{FD6864} }}                                                                                  \\
		\multirow{-2}{*}{\textbf{Model}} & \multicolumn{1}{l}{\multirow{-2}{*}{\textbf{h-step}}} & \multirow{-2}{*}{\textbf{\begin{tabular}[c]{@{}c@{}}Actual \\ value\end{tabular}}} & \multicolumn{1}{l}{\multirow{-2}{*}{\textbf{Mean}}} & \multicolumn{1}{l}{\multirow{-2}{*}{\textbf{Median}}} & \multicolumn{1}{l}{\multirow{-2}{*}{\textbf{Mode}}} & \multicolumn{1}{l}{\multirow{-2}{*}{{\textbf{\begin{tabular}[c]{@{}l@{}}HPP\\ Interval\end{tabular}}}}} \\ \hline
		NoGeAR(1)                        & 1                                                     & 0                                                                                  & 0.55                                                & 1                                                     & 0                                                   & { $\{0,\dots,4\}$}                                                                                       \\
		& 2                                                     & 3                                                                                  & 0.58                                                & 1                                                     & 0                                                   & { $\{0,\dots,4\}$}                                                                                       \\
		NGINAR                           & 1                                                     & 0                                                                                  & 1.01                                                & 0                                                     & 0                                                   & { $\{0,\dots,4\}$}                                                                                       \\
		& 2                                                     & 3                                                                                  & 1.15                                                & 1                                                     & 0                                                   & { $\{0,\dots,4\}$}                                                                                       \\
		NBINAR                           & 1                                                     & 0                                                                                  & 0.26                                                & 0                                                     & 0                                                   & { $\{0,1\}$}                                                                                             \\
		& 2                                                     & 3                                                                                  & 0.30                                                & 3                                                     & 0                                                   & { $\{0,\dots,7\}$}                                                                                       \\
		GINAR                            & 1                                                     & 0                                                                                  & 1.03                                                & 0                                                     & 0                                                   & { $\{0,\dots,4\}$}                                                                                       \\
		& 2                                                     & 3                                                                                  & 1.09                                                & 1                                                     & 0                                                   & { $\{0,\dots,4\}$}                                                                                       \\
		PINAR                            & 1                                                     & 0                                                                                  & 1.05                                                & 1                                                     & 1                                                   & { $\{0,\dots,3\}$}                                                                                       \\
		& 2                                                     & 3                                                                                  & 1.11                                                & 1                                                     & 1                                                   & { $\{0,\dots,3\}$}                                                                                       \\ \hline
	\end{tabular}
		}
		\label{last}
	\end{table}

	From \textcolor{blue}{Figures} \ref{fig:6} and \ref{fig:2}, significant lags of order greater than or equal to 1 could be observed in the ACF plots which leads to question the appropriateness of INAR(1) model for the respective data. To verify the validity of an INAR(1) model to such data, \cite{forughi2022portmanteau} developed Portmanteau tests in which the null hypothesis states that an INAR model of order 1 is adequate for the data versus an INAR model of order $p$, with $p > 1$ is adequate for the data. In the same work, they have established that the TeXpert data can be well explained by an INAR(1) model. The p-values corresponding to the Portmanteau tests on the Barbados data are presented in \textcolor{blue}{Table} \ref{port}. The p- values affirm the INAR(1) choice for the data.
	
		\begin{figure}[H]
		\centering
		\includegraphics[scale=0.5]{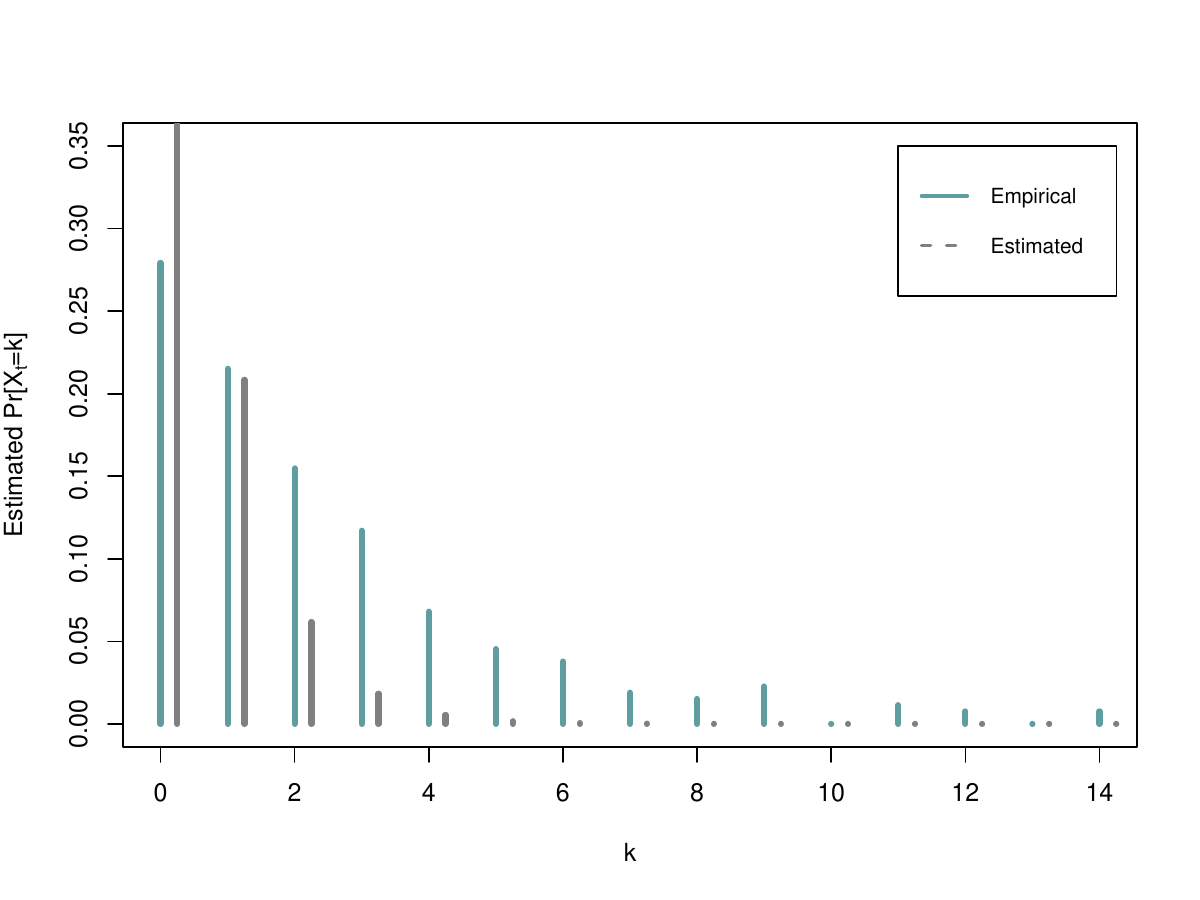}
		\caption{Plot of empirical frequencies( blue) and a geometric fit( gray) for CWß TeXpert downloads data}
		\label{fig:3}
	\end{figure}
		
	\begin{table}[H]
		\center
		\caption{Saddlepoint Approximation Estimates, AIC, BIC and  AIC$_c$ values for various models fitted to CWß TeXpert downloads data }
		\renewcommand{\arraystretch}{1.5}
		\scalebox{0.8}{%
			\begin{tabular}{@{}llccccl@{}}
				\toprule
				\multirow{2}{*}{\textbf{Model}} & \multicolumn{3}{c}{\textbf{Parameter}}                                                                              & \multirow{2}{*}{\textbf{AIC}}                & \multicolumn{1}{l}{\multirow{2}{*}{\textbf{BIC}}} & \multirow{2}{*}{\textbf{AIC$_c$}} \\ \cmidrule(lr){2-4}
				& 1                                           & 2                                           & 3                       &                                              & \multicolumn{1}{l}{}                              &                                   \\ \midrule
				NoGeAR(1)                       & \multirow{2}{*}{0.8649}                     & \multirow{2}{*}{0.5378}                     & \multirow{2}{*}{0.6993} & \multirow{2}{*}{1036.81}                     & \multirow{2}{*}{1043.97}                          & \multirow{2}{*}{1036.86}          \\
				($\alpha, \beta, \theta$)       &                                             &                                             &                         &                                              &                                                   &                                   \\
				NGINAR                          & \multirow{2}{*}{1.3789}                     & \multirow{2}{*}{0.2410}                     & \multirow{2}{*}{}       & \multirow{2}{*}{1063.30}                     & \multirow{2}{*}{1074.00}                          & \multirow{2}{*}{1063.40}          \\
				($\mu, \alpha$)                 &                                             &                                             &                         &                                              &                                                   &                                   \\
				NBINAR                          & \multirow{2}{*}{2.0999}                     & \multirow{2}{*}{0.9002}                     & \multirow{2}{*}{0.1000} & \multicolumn{1}{l}{\multirow{2}{*}{1069.20}} & \multirow{2}{*}{1080.03}                          & \multirow{2}{*}{1069.30}          \\
				($n, p, \alpha$)                &                                             &                                             &                         & \multicolumn{1}{l}{}                         &                                                   &                                   \\
				GINAR                           & \multicolumn{1}{c}{\multirow{2}{*}{0.6962}} & \multirow{2}{*}{0.1453}                     & \multirow{2}{*}{}       & \multirow{2}{*}{1069.81}                     & \multirow{2}{*}{1076.97}                          & \multirow{2}{*}{1069.85}          \\
				($p, \alpha$)                   & \multicolumn{1}{c}{}                        &                                             &                         &                                              &                                                   &                                   \\
				PINAR                           & \multirow{2}{*}{2.3384}                     & \multicolumn{1}{l}{\multirow{2}{*}{0.1683}} & \multirow{2}{*}{}       & \multirow{2}{*}{1259.19}                     & \multirow{2}{*}{1266.35}                          & \multirow{2}{*}{1259.24}          \\
				($\lambda, \alpha$)             &                                             & \multicolumn{1}{l}{}                        &                         &                                              &                                                   &                                   \\ \bottomrule
			\end{tabular}
		}
		\label{tab6}
	\end{table}
		\section{Conclusion}\label{conc}	
	
	\noindent In this paper, we have presented the problem of coherent forecasting within the NoGeAR(1) model framework. By utilizing the conditional distribution and MC approximation, we generate coherent forecasts. A comprehensive simulation study has been conducted to assess the model's forecasting performance with its special case - NGINAR model. The proposed coherent forecasting methodology is then demonstrated through the analysis of two real datasets. Model adequacy is confirmed using Pearson's residuals and PIT histogram. The study reveals that NoGeAR(1) performs either better or close to other overdispersed count time series models that fit the data well. 

	\begin{table}[H]
		\center
		\caption{Point forecasts, HPP intervals of $95\%$ coverage probability for Downloads data}
		\renewcommand{\arraystretch}{1.7}
		\scalebox{0.9}{%
			\begin{tblr}{
					column{3} = {c},
					column{7} = {l},
					cell{2}{2} = {c},
					cell{2}{4} = {c},
					cell{2}{5} = {c},
					cell{2}{6} = {c},
					cell{2}{7} = {c},
					cell{3}{2} = {c},
					cell{3}{4} = {c},
					cell{3}{5} = {c},
					cell{3}{6} = {c},
					cell{3}{7} = {c},
					cell{4}{2} = {c},
					cell{4}{4} = {c},
					cell{4}{5} = {c},
					cell{4}{6} = {c},
					cell{4}{7} = {c},
					cell{5}{2} = {c},
					cell{5}{4} = {c},
					cell{5}{5} = {c},
					cell{5}{6} = {c},
					cell{5}{7} = {c},
					cell{6}{2} = {c},
					cell{6}{4} = {c},
					cell{6}{5} = {c},
					cell{6}{6} = {c},
					cell{6}{7} = {c},
					cell{7}{2} = {c},
					cell{7}{4} = {c},
					cell{7}{5} = {c},
					cell{7}{6} = {c},
					cell{7}{7} = {c},
					cell{8}{2} = {c},
					cell{8}{4} = {c},
					cell{8}{5} = {c},
					cell{8}{6} = {c},
					cell{8}{7} = {c},
					cell{9}{2} = {c},
					cell{9}{4} = {c},
					cell{9}{5} = {c},
					cell{9}{6} = {c},
					cell{9}{7} = {c},
					cell{10}{2} = {c},
					cell{10}{4} = {c},
					cell{10}{5} = {c},
					cell{10}{6} = {c},
					cell{10}{7} = {c},
					cell{11}{2} = {c},
					cell{11}{4} = {c},
					cell{11}{5} = {c},
					cell{11}{6} = {c},
					cell{11}{7} = {c},
					hline{1-2,12} = {-}{},
				}
				\textbf{Model} & \textbf{h-step} & {\textbf{Actual }\\\textbf{ value}} & \textbf{Mean} & \textbf{Median} & \textbf{Mode} & {\textbf{HPP}\\\textbf{Interval}} \\
				NoGeAR(1)      & 1               & 4                                   & 1.77          & 3               & 1             & $\{0,\dots,7\}$                    \\
				& 2               & 7                                   & 1.47          & 1               & 0             & $\{0,\dots,7\}$                    \\
				NGINAR         & 1               & 4                                   & 1.57          & 2               & 0             & $\{0,\dots,7\}$                    \\
				& 2               & 7                                   & 1.07          & 1               & 0             & $\{0,\dots,7\}$                    \\
				NBINAR         & 1               & 4                                   & 2.74          & 10              & 7             & $\{0,\dots,14\}$                   \\
				& 2               & 7                                   & 2.51          & 0               & 0             & $\{0,\dots,14\}$                   \\
				GINAR          & 1               & 4                                   & 2.39          & 2               & 0             & $\{0,\dots,7\}$                    \\
				& 2               & 7                                   & 2.31          & 1               & 0             & $\{0,\dots,9\}$                    \\
				PINAR          & 1               & 4                                   & 2.45          & 2               & 2             & $\{0,\dots,4\}$                    \\
				& 2               & 7                                   & 2.35          & 2               & 2             & $\{0,\dots,7\}$                    
			\end{tblr}}
		\label{tab7}
	\end{table}

	\section*{Declaration of interest}
	\noindent  No potential conflict of interest was reported by the authors.
	
	\section*{Acknowledgement}
	The authors thank the referees for their constructive comments on an earlier draft of the paper. Divya Kuttenchalil Andrews wishes to thank Cochin University of Science and Technology, India, for the financial support.

	\bibliography{§}
	\bibliographystyle{apalike}
	\begin{appendices}
		\section{Proof of Theorem 3.1}\label{appA}
		 In this Appendix, we present the proof of \textcolor{blue}{Theorem} \ref{thm1}. As mentioned in the article, we make use of (\ref{eq:for2}) i.e.,
			\begin{equation}
				Pr[X_{t+2}=x|X_t = y] = \sum_{k=0}^\infty Pr[X_{t+2}=x|X_{t+1} = k]Pr[X_{t+1}=k|X_t = y].\nonumber
			\end{equation} 
			Using the one-step ahead transition probabilities specified in (\ref{eqn: eq a}), we have the respective component probabilities in (\ref{eq:for2}) as:
			\begin{equation}
				\label{eqa1}
				Pr[X_{t+2} = x| X_{t+1} = k] =\begin{cases}
					1 -\alpha \theta , & \textrm{if } x=0,\;k=0\\
					Pr[\varepsilon_{t+1} =x], & \textrm{if } x \geq 1,\;k=0\\
					\alpha ^k (1 -\alpha \theta) , & \textrm{if } x=0,\;k\geq 1\\
					\alpha ^k Pr[\varepsilon_{t+1} = x]\\
					+ \sum_{m = 1}^{x}\sum_{j = 1}^{min(m,k)} Pr[\varepsilon_{t+1} =x-m]\left(\begin{array}{l}
						k \\
						j
					\end{array}\right)\\   \times \left(\begin{array}{c}
						m-1 \\
						j-1
					\end{array}\right) [(1-\alpha)(1-\beta)]^j \alpha ^{k-j}\beta^{m-j}, &\textrm{if }x \geq 1,\;k\geq 1.
				\end{cases}  
			\end{equation}
			and
			\begin{equation}
				\label{eqa2}
				Pr[X_{t+1} = k| X_{t} = y] =\begin{cases}
					1 -\alpha \theta , & \textrm{if } k=0,\;y=0\\
					Pr[\varepsilon_t =k], & \textrm{if } k \geq 1,\;y=0\\
					\alpha ^y (1 -\alpha \theta) , & \textrm{if } k=0,\;y\geq 1\\
					\alpha ^y Pr[\varepsilon_t = k]\\
					+ \sum_{l = 1}^{k}\sum_{p = 1}^{min(l,y)} Pr[\varepsilon_t =k-l]\left(\begin{array}{l}
						y \\
						p
					\end{array}\right)\\   \times \left(\begin{array}{c}
						l-1 \\
						p-1
					\end{array}\right) [(1-\alpha)(1-\beta)]^p \alpha ^{y-p}\beta^{l-p}, &\textrm{if }k \geq 1,\;y\geq 1.
				\end{cases} 
			\end{equation}\\
			\textit{Case (i)} $k=0$\\
			In this case, we have the following possible combinations:
			\begin{enumerate}
				\item[(a)] $(x=0,k=0) \times (k=0,y=0) \Rightarrow (1-\alpha\theta)(1-\alpha\theta) \equiv (1-\alpha\theta)^2$.
				\item[(b)] $(x\geq 1,k=0) \times (k=0,y=0) \Rightarrow Pr[\varepsilon_{t+1} = x](1-\alpha\theta) \equiv (1-\alpha\theta)Pr[\varepsilon_{t} = x]$.
				\item[(c)] $(x=0,k=0) \times (k=0,y\geq 1) \Rightarrow (1-\alpha\theta)\alpha^y(1-\alpha\theta) \equiv \alpha^y(1-\alpha\theta)^2$.
				\small
				\item[(d)]  $(x\geq 1,k=0) \times (k=0,y\geq 1) \Rightarrow Pr[\varepsilon_{t+1} = x]\alpha^y(1-\alpha\theta) \equiv \alpha^y (1-\alpha\theta)Pr[\varepsilon_{t} = x]$.\\
			\end{enumerate}
			\normalsize 
			\textit{Case (ii)} $k\geq 1$\\
			As seen previously, we have the following possibilities:
			\small
			\begin{enumerate}
				\item[(a)] $(x=0,k\geq 1) \times (k\geq 1,y=0) \Rightarrow \alpha^k (1-\alpha\theta)Pr[\varepsilon_{t} = k]$.
				\item[(b)] $(x\geq 1,k\geq 1) \times (k\geq 1,y=0) \Rightarrow Pr[\varepsilon_{t} = x]\alpha^kPr[\varepsilon_{t} = k] + $\\$  \sum_{m = 1}^{x}\sum_{j = 1}^{min(m,k)}Pr[\varepsilon_{t} =k] Pr[\varepsilon_{t} =x-m]\left(\begin{array}{l}
					k \\
					j
				\end{array}\right) \left(\begin{array}{c}
					m-1 \\
					j-1
				\end{array}\right) [(1-\alpha)(1-\beta)]^j \alpha ^{k-j}\beta^{m-j}.$
				\item[(c)] $(x=0,k\geq 1) \times (k\geq 1,y\geq 1) \Rightarrow \alpha^{k+y} (1-\alpha\theta)Pr[\varepsilon_{t} = k] + $\\
				$\sum_{l = 1}^{k}\sum_{p = 1}^{min(l,y)} Pr[\varepsilon_t =k-l]\left(\begin{array}{l}
					y \\
					p
				\end{array}\right)\left(\begin{array}{c}
					l-1 \\
					p-1
				\end{array}\right) [(1-\alpha)(1-\beta)]^p \alpha ^{k+y-p}\beta^{l-p}(1-\alpha\theta)$.
				\item[(d)]  $(x\geq 1,k\geq 1) \times (k\geq 1,y\geq 1) \Rightarrow \\
				\Big\{\alpha ^k Pr[\varepsilon_{t+1} = x] + \sum_{m = 1}^{x}\sum_{j = 1}^{min(m,k)} Pr[\varepsilon_{t+1} =x-m]\left(\begin{array}{l}
					k \\
					j
				\end{array}\right) \left(\begin{array}{c}
					m-1 \\
					j-1
				\end{array}\right)\\  \times [(1-\alpha)(1-\beta)]^j \alpha ^{k-j}\beta^{m-j}\Big\} \Big\{\alpha ^y Pr[\varepsilon_t = k]
				+ \sum_{l = 1}^{k}\sum_{p = 1}^{min(l,y)} Pr[\varepsilon_t =k-l]\\ \times \left(\begin{array}{l}
					y \\
					p
				\end{array}\right) \left(\begin{array}{c}
					l-1 \\
					p-1
				\end{array}\right) [(1-\alpha)(1-\beta)]^p \alpha ^{y-p}\beta^{l-p}\Big\}\\
				\equiv \Big\{ \alpha ^{k+y} Pr[\varepsilon_{t} = x] Pr[\varepsilon_{t} = k] + \sum_{l = 1}^{k}\sum_{p = 1}^{min(l,y)} Pr[\varepsilon_t = x]Pr[\varepsilon_t =k-l] \left(\begin{array}{l}
					y \\
					p
				\end{array}\right) \left(\begin{array}{c}
					l-1 \\
					p-1
				\end{array}\right) \\ \times [(1-\alpha)(1-\beta)]^p \alpha ^{k+y-p}\beta^{l-p}\Big\} + \Big\{ \sum_{m = 1}^{x}\sum_{j = 1}^{min(m,k)} Pr[\varepsilon_{t} =k]Pr[\varepsilon_{t} =x-m]\\ \times\left(\begin{array}{l}
					k \\
					j
				\end{array}\right) \left(\begin{array}{c}
					m-1 \\
					j-1
				\end{array}\right) [(1-\alpha)(1-\beta)]^j \alpha ^{k+y-j}\beta^{m-j} \Big\} + \Big\{ \sum_{m = 1}^{x}\sum_{j = 1}^{min(m,k)} Pr[\varepsilon_{t} =x-m]\left(\begin{array}{l}
					k \\
					j
				\end{array}\right) \left(\begin{array}{c}
					m-1 \\
					j-1
				\end{array}\right) [(1-\alpha)(1-\beta)]^j \alpha ^{k-j}\beta^{m-j}\Big\} \times \Big\{\sum_{l = 1}^{k}\sum_{p = 1}^{min(l,y)} Pr[\varepsilon_t =k-l]\left(\begin{array}{l}
					y \\
					p
				\end{array}\right)\left(\begin{array}{c}
					l-1 \\
					p-1
				\end{array}\right) [(1-\alpha)(1-\beta)]^p \alpha ^{y-p}\beta^{l-p} \Big\} $. 
			\end{enumerate}
			\normalsize
			Thus, using (\ref{eq:for2}), (\ref{eqa1}), (\ref{eqa2}) and substituting the respective probabilities for the cases above, we can arrive at (\ref{eqn: eq a2}). Hence, the proof.
	\end{appendices}
\end{document}